\documentclass[amsmath,amssymb,aps,prd,11pt,tightenlines,superscriptaddress,nofootinbib,preprintnumbers,notitlepage]{revtex4-1}

\input{header}
\newcommand{\PRE}[1]{{#1}}
\usepackage{xspace}
\newcommand{\pythiaF}{\texttt{PYTHIAforward}\xspace}
\usepackage{orcidlink}

\begin{document}

\preprint{CERN-EP-2024-036}

\title{{\Large Neutrino Rate Predictions for FASER} \vspace*{0.15in} \\    
FASER Collaboration}

\author{Roshan Mammen Abraham\,\orcidlink{0000-0003-4678-3808}}
\affiliation{Department of Physics and Astronomy, University of California, Irvine, CA 92697-4575, USA}

\author{John Anders\,\orcidlink{0000-0002-1846-0262}}
\affiliation{CERN, CH-1211 Geneva 23, Switzerland}

\author{Claire Antel\,\orcidlink{0000-0001-9683-0890}}
\affiliation{D\'epartement de Physique Nucl\'eaire et Corpusculaire, University of Geneva, CH-1211 Geneva 4, Switzerland}

\author{Akitaka Ariga\,\orcidlink{0000-0002-6832-2466}}
\affiliation{Albert Einstein Center for Fundamental Physics, Laboratory for High Energy Physics, University of Bern, Sidlerstrasse 5, CH-3012 Bern, Switzerland}
\affiliation{Department of Physics, Chiba University, 1-33 Yayoi-cho Inage-ku, 263-8522 Chiba, Japan}

\author{Tomoko Ariga\,\orcidlink{0000-0001-9880-3562}}
\affiliation{Kyushu University, Nishi-ku, 819-0395 Fukuoka, Japan}

\author{Jeremy Atkinson\,\orcidlink{0009-0003-3287-2196}} 
\affiliation{Albert Einstein Center for Fundamental Physics, Laboratory for High Energy Physics, University of Bern, Sidlerstrasse 5, CH-3012 Bern, Switzerland}

\author{Florian~U.~Bernlochner\,\orcidlink{0000-0001-8153-2719}} 
\affiliation{Universit\"at Bonn, Regina-Pacis-Weg 3, D-53113 Bonn, Germany}

\author{Tobias Boeckh\,\orcidlink{0009-0000-7721-2114}} 
\affiliation{Universit\"at Bonn, Regina-Pacis-Weg 3, D-53113 Bonn, Germany}

\author{Jamie Boyd\,\orcidlink{0000-0001-7360-0726}} 
\affiliation{CERN, CH-1211 Geneva 23, Switzerland}

\author{Lydia Brenner\,\orcidlink{0000-0001-5350-7081}}
\affiliation{Nikhef National Institute for Subatomic Physics, Science Park 105, 1098 XG Amsterdam, Netherlands}

\author{Angela Burger\,\orcidlink{0000-0003-0685-4122}}
\affiliation{CERN, CH-1211 Geneva 23, Switzerland}

\author{Franck Cadoux} 
\affiliation{D\'epartement de Physique Nucl\'eaire et Corpusculaire, University of Geneva, CH-1211 Geneva 4, Switzerland}

\author{Roberto Cardella\,\orcidlink{0000-0002-3117-7277}} 
\affiliation{D\'epartement de Physique Nucl\'eaire et Corpusculaire, University of Geneva, CH-1211 Geneva 4, Switzerland}

\author{David~W.~Casper\,\orcidlink{0000-0002-7618-1683}} 
\affiliation{Department of Physics and Astronomy, University of California, Irvine, CA 92697-4575, USA}

\author{Charlotte Cavanagh\,\orcidlink{0009-0001-1146-5247}} \affiliation{University of Liverpool, Liverpool L69 3BX, United Kingdom}

\author{Xin Chen\,\orcidlink{0000-0003-4027-3305}} 
\affiliation{Department of Physics, Tsinghua University, Beijing, China}

\author{Andrea Coccaro\,\orcidlink{0000-0003-2368-4559}} 
\affiliation{INFN Sezione di Genova, Via Dodecaneso, 33--16146, Genova, Italy}

\author{Stephane D\'{e}bieux} 
\affiliation{D\'epartement de Physique Nucl\'eaire et Corpusculaire, University of Geneva, CH-1211 Geneva 4, Switzerland}

\author{Monica D’Onofrio\,\orcidlink{0000-0003-2408-5099}} 
\affiliation{University of Liverpool, Liverpool L69 3BX, United Kingdom}

\author{Ansh Desai\,\orcidlink{0000-0002-5447-8304}} 
\affiliation{University of Oregon, Eugene, OR 97403, USA}

\author{Sergey Dmitrievsky\,\orcidlink{0000-0003-4247-8697}} 
\affiliation{Affiliated with an international laboratory covered by a cooperation agreement with CERN.}

\author{Sinead Eley\,\orcidlink{0009-0001-1320-2889}}
\affiliation{University of Liverpool, Liverpool L69 3BX, United Kingdom}

\author{Yannick Favre} 
\affiliation{D\'epartement de Physique Nucl\'eaire et Corpusculaire, University of Geneva, CH-1211 Geneva 4, Switzerland}

\author{Deion Fellers\,\orcidlink{0000-0002-0731-9562}} 
\affiliation{University of Oregon, Eugene, OR 97403, USA}

\author{Jonathan~L.~Feng\,\orcidlink{0000-0002-7713-2138}} 
\affiliation{Department of Physics and Astronomy, University of California, Irvine, CA 92697-4575, USA}

\author{Carlo Alberto Fenoglio\,\orcidlink{0009-0007-7567-8763}}  
\affiliation{D\'epartement de Physique Nucl\'eaire et Corpusculaire, University of Geneva, CH-1211 Geneva 4, Switzerland}

\author{Didier Ferrere\,\orcidlink{0000-0002-5687-9240}} 
\affiliation{D\'epartement de Physique Nucl\'eaire et Corpusculaire, University of Geneva, CH-1211 Geneva 4, Switzerland}

\author{Max Fieg\,\orcidlink{0000-0002-7027-6921}} 
\affiliation{Department of Physics and Astronomy, University of California, Irvine, CA 92697-4575, USA}

\author{Wissal Filali\,\orcidlink{0009-0008-6961-2335}} 
\affiliation{Universit\"at Bonn, Regina-Pacis-Weg 3, D-53113 Bonn, Germany}

\author{Stephen Gibson\,\orcidlink{0000-0002-1236-9249}} 
\affiliation{Royal Holloway, University of London, Egham, TW20 0EX, United Kingdom}

\author{Sergio Gonzalez-Sevilla\,\orcidlink{0000-0003-4458-9403}} 
\affiliation{D\'epartement de Physique Nucl\'eaire et Corpusculaire, University of Geneva, CH-1211 Geneva 4, Switzerland}

\author{Yuri Gornushkin\,\orcidlink{0000-0003-3524-4032}} 
\affiliation{Affiliated with an international laboratory covered by a cooperation agreement with CERN.}

\author{Carl Gwilliam\,\orcidlink{0000-0002-9401-5304}} 
\affiliation{University of Liverpool, Liverpool L69 3BX, United Kingdom}

\author{Daiki Hayakawa\,\orcidlink{0000-0003-4253-4484}} 
\affiliation{Department of Physics, Chiba University, 1-33 Yayoi-cho Inage-ku, 263-8522 Chiba, Japan}

\author{Shih-Chieh Hsu\,\orcidlink{0000-0001-6214-8500}} 
\affiliation{Department of Physics, University of Washington, PO Box 351560, Seattle, WA 98195-1460, USA}

\author{Zhen Hu\,\orcidlink{0000-0001-8209-4343}} 
\affiliation{Department of Physics, Tsinghua University, Beijing, China}

\author{Giuseppe Iacobucci\,\orcidlink{0000-0001-9965-5442}} 
\affiliation{D\'epartement de Physique Nucl\'eaire et Corpusculaire, University of Geneva, CH-1211 Geneva 4, Switzerland}

\author{Tomohiro Inada\,\orcidlink{0000-0002-6923-9314}} 
\affiliation{CERN, CH-1211 Geneva 23, Switzerland}

\author{Luca Iodice\,\orcidlink{0000-0002-3516-7121}} 
\affiliation{D\'epartement de Physique Nucl\'eaire et Corpusculaire, University of Geneva, CH-1211 Geneva 4, Switzerland}

\author{Sune Jakobsen\,\orcidlink{0000-0002-6564-040X}} 
\affiliation{CERN, CH-1211 Geneva 23, Switzerland}

\author{Hans Joos\,\orcidlink{0000-0003-4313-4255}} 
\affiliation{CERN, CH-1211 Geneva 23, Switzerland}
\affiliation{II.~Physikalisches Institut, Universität Göttingen, Göttingen, Germany}

\author{Enrique Kajomovitz\,\orcidlink{0000-0002-8464-1790}} 
\affiliation{Department of Physics and Astronomy, Technion---Israel Institute of Technology, Haifa 32000, Israel}

\author{Hiroaki Kawahara\,\orcidlink{0009-0007-5657-9954}}
\affiliation{Kyushu University, Nishi-ku, 819-0395 Fukuoka, Japan}

\author{Alex Keyken\,\orcidlink{0009-0001-4886-2924}}
\affiliation{Royal Holloway, University of London, Egham, TW20 0EX, United Kingdom}

\author{Felix Kling\,\orcidlink{0000-0002-3100-6144}} 
\affiliation{Deutsches Elektronen-Synchrotron DESY, Notkestr. 85, 22607 Hamburg, Germany}

\author{Daniela Köck\,\orcidlink{0000-0002-9090-5502}}     
\affiliation{University of Oregon, Eugene, OR 97403, USA}

\author{Pantelis Kontaxakis\,\orcidlink{0000-0002-4860-5979}} 
\affiliation{D\'epartement de Physique Nucl\'eaire et Corpusculaire, University of Geneva, CH-1211 Geneva 4, Switzerland}

\author{Umut Kose\,\orcidlink{0000-0001-5380-9354}} 
\affiliation{ETH Zurich, 8092 Zurich, Switzerland}

\author{Rafaella Kotitsa\,\orcidlink{0000-0002-7886-2685}} 
\affiliation{CERN, CH-1211 Geneva 23, Switzerland}

\author{Susanne Kuehn\,\orcidlink{0000-0001-5270-0920}} 
\affiliation{CERN, CH-1211 Geneva 23, Switzerland}

\author{Thanushan Kugathasan\,\orcidlink{0000-0003-4631-5019}} 
\affiliation{D\'epartement de Physique Nucl\'eaire et Corpusculaire, University of Geneva, CH-1211 Geneva 4, Switzerland}

\author{Helena Lefebvre\,\orcidlink{0000-0002-7394-2408}} 
\affiliation{Royal Holloway, University of London, Egham, TW20 0EX, United Kingdom}

\author{Lorne Levinson\,\orcidlink{0000-0003-4679-0485}} 
\affiliation{Department of Particle Physics and Astrophysics, Weizmann Institute of Science, Rehovot 76100, Israel}

\author{Ke Li\,\orcidlink{0000-0002-2545-0329}} 
\affiliation{Department of Physics, University of Washington, PO Box 351560, Seattle, WA 98195-1460, USA}

\author{Jinfeng Liu\,\orcidlink{0000-0001-6827-1729}}
\affiliation{Department of Physics, Tsinghua University, Beijing, China}

\author{Margaret S.~Lutz\,\orcidlink{0000-0003-4515-0224}}
\affiliation{CERN, CH-1211 Geneva 23, Switzerland}

\author{Jack MacDonald\,\orcidlink{0000-0002-3150-3124}}    
\affiliation{Institut f\"ur Physik, Universität Mainz, Mainz, Germany}

\author{Chiara Magliocca\,\orcidlink{0009-0009-4927-9253}} 
\affiliation{D\'epartement de Physique Nucl\'eaire et Corpusculaire, University of Geneva, CH-1211 Geneva 4, Switzerland}

\author{Fulvio Martinelli\,\orcidlink{0000-0003-4221-5862}} 
\affiliation{D\'epartement de Physique Nucl\'eaire et Corpusculaire, University of Geneva, CH-1211 Geneva 4, Switzerland}

\author{Lawson McCoy\,\orcidlink{0009-0009-2741-3220}} 
\affiliation{Department of Physics and Astronomy, University of California, Irvine, CA 92697-4575, USA}

\author{Josh McFayden\,\orcidlink{0000-0001-9273-2564}} 
\affiliation{Department of Physics \& Astronomy, University of Sussex, Sussex House, Falmer, Brighton, BN1 9RH, United Kingdom}

\author{Andrea Pizarro Medina\,\orcidlink{0000-0002-1024-5605}} 
\affiliation{D\'epartement de Physique Nucl\'eaire et Corpusculaire, University of Geneva, CH-1211 Geneva 4, Switzerland}

\author{Matteo Milanesio\,\orcidlink{0000-0001-8778-9638}} 
\affiliation{D\'epartement de Physique Nucl\'eaire et Corpusculaire, University of Geneva, CH-1211 Geneva 4, Switzerland}

\author{Théo Moretti\,\orcidlink{0000-0001-7065-1923}} 
\affiliation{D\'epartement de Physique Nucl\'eaire et Corpusculaire, University of Geneva, CH-1211 Geneva 4, Switzerland}

\author{Magdalena Munker\,\orcidlink{0000-0003-2775-3291}} 
\affiliation{D\'epartement de Physique Nucl\'eaire et Corpusculaire, University of Geneva, CH-1211 Geneva 4, Switzerland}

\author{Mitsuhiro Nakamura} 
\affiliation{Nagoya University, Furo-cho, Chikusa-ku, Nagoya 464-8602, Japan}

\author{Toshiyuki Nakano} 
\affiliation{Nagoya University, Furo-cho, Chikusa-ku, Nagoya 464-8602, Japan}

\author{Friedemann Neuhaus\,\orcidlink{0000-0002-3819-2453}} 
\affiliation{Institut f\"ur Physik, Universität Mainz, Mainz, Germany}

\author{Laurie Nevay\,\orcidlink{0000-0001-7225-9327}} 
\affiliation{CERN, CH-1211 Geneva 23, Switzerland}

\author{Ken Ohashi\,\orcidlink{0009-0000-9494-8457}}
\affiliation{Albert Einstein Center for Fundamental Physics, Laboratory for High Energy Physics, University of Bern, Sidlerstrasse 5, CH-3012 Bern, Switzerland}

\author{Hidetoshi Otono\,\orcidlink{0000-0003-0760-5988}} 
\affiliation{Kyushu University, Nishi-ku, 819-0395 Fukuoka, Japan}

\author{Hao Pang\,\orcidlink{0000-0002-1946-1769}} 
\affiliation{Department of Physics, Tsinghua University, Beijing, China}

\author{Lorenzo Paolozzi\,\orcidlink{0000-0002-9281-1972}} 
\affiliation{D\'epartement de Physique Nucl\'eaire et Corpusculaire, University of Geneva, CH-1211 Geneva 4, Switzerland}
\affiliation{CERN, CH-1211 Geneva 23, Switzerland}

\author{Brian Petersen\,\orcidlink{0000-0002-7380-6123}} 
\affiliation{CERN, CH-1211 Geneva 23, Switzerland}

\author{Markus Prim\,\orcidlink{0000-0002-1407-7450}} 
\affiliation{Universit\"at Bonn, Regina-Pacis-Weg 3, D-53113 Bonn, Germany}

\author{Michaela Queitsch-Maitland\,\orcidlink{0000-0003-4643-515X}} 
\affiliation{University of Manchester, School of Physics and Astronomy, Schuster Building, Oxford Rd, Manchester M13 9PL, United Kingdom}

\author{Hiroki Rokujo\,\orcidlink{0000-0002-3502-493X}}
\affiliation{Nagoya University, Furo-cho, Chikusa-ku, Nagoya 464-8602, Japan}

\author{Elisa Ruiz-Choliz\,\orcidlink{0000-0002-2417-7121}} 
\affiliation{Institut f\"ur Physik, Universität Mainz, Mainz, Germany}

\author{Andr\'e Rubbia\,\orcidlink{0000-0002-5747-1001}} 
\affiliation{ETH Zurich, 8092 Zurich, Switzerland}

\author{Jorge Sabater-Iglesias\,\orcidlink{0000-0003-2328-1952}} 
\affiliation{D\'epartement de Physique Nucl\'eaire et Corpusculaire, University of Geneva, CH-1211 Geneva 4, Switzerland}

\author{Osamu Sato\,\orcidlink{0000-0002-6307-7019}} 
\affiliation{Nagoya University, Furo-cho, Chikusa-ku, Nagoya 464-8602, Japan}

\author{Paola Scampoli\,\orcidlink{0000-0001-7500-2535}} 
\affiliation{Albert Einstein Center for Fundamental Physics, Laboratory for High Energy Physics, University of Bern, Sidlerstrasse 5, CH-3012 Bern, Switzerland}
\affiliation{Dipartimento di Fisica ``Ettore Pancini'', Universit\`a di Napoli Federico II, Complesso Universitario di Monte S. Angelo, I-80126 Napoli, Italy}

\author{Kristof Schmieden\,\orcidlink{0000-0003-1978-4928}} 
\affiliation{Institut f\"ur Physik, Universität Mainz, Mainz, Germany}

\author{Matthias Schott\,\orcidlink{0000-0002-4235-7265}} 
\affiliation{Institut f\"ur Physik, Universität Mainz, Mainz, Germany}

\author{Anna Sfyrla\,\orcidlink{0000-0002-3003-9905}} 
\affiliation{D\'epartement de Physique Nucl\'eaire et Corpusculaire, University of Geneva, CH-1211 Geneva 4, Switzerland}

\author{Mansoora Shamim\,\orcidlink{0009-0002-3986-399X}}
\affiliation{CERN, CH-1211 Geneva 23, Switzerland}

\author{Savannah Shively\,\orcidlink{0000-0002-4691-3767}} 
\affiliation{Department of Physics and Astronomy, University of California, Irvine, CA 92697-4575, USA}

\author{Yosuke Takubo\,\orcidlink{0000-0002-3143-8510}} 
\affiliation{Institute of Particle and Nuclear Studies, KEK, Oho 1-1, Tsukuba, Ibaraki 305-0801, Japan}

\author{Noshin Tarannum\,\orcidlink{0000-0002-3246-2686}} 
\affiliation{D\'epartement de Physique Nucl\'eaire et Corpusculaire, University of Geneva, CH-1211 Geneva 4, Switzerland}

\author{Ondrej Theiner\,\orcidlink{0000-0002-6558-7311}} 
\affiliation{D\'epartement de Physique Nucl\'eaire et Corpusculaire, University of Geneva, CH-1211 Geneva 4, Switzerland}

\author{Eric Torrence\,\orcidlink{0000-0003-2911-8910}} 
\affiliation{University of Oregon, Eugene, OR 97403, USA}

\author{Svetlana Vasina\,\orcidlink{0000-0003-2775-5721}} 
\affiliation{Affiliated with an international laboratory covered by a cooperation agreement with CERN.}

\author{Benedikt Vormwald\,\orcidlink{0000-0003-2607-7287}} 
\affiliation{CERN, CH-1211 Geneva 23, Switzerland}

\author{Di Wang\,\orcidlink{0000-0002-0050-612X}} 
\affiliation{Department of Physics, Tsinghua University, Beijing, China}

\author{Yuxiao Wang\,\orcidlink{0009-0004-1228-9849}} 
\affiliation{Department of Physics, Tsinghua University, Beijing, China}

\author{Eli Welch\,\orcidlink{0000-0001-6336-2912}} 
\affiliation{Department of Physics and Astronomy, University of California, Irvine, CA 92697-4575, USA}

\author{Samuel Zahorec\,\orcidlink{0009-0000-9729-0611}}
\affiliation{CERN, CH-1211 Geneva 23, Switzerland}
\affiliation{Charles University, Faculty of Mathematics and Physics, Prague; Czech Republic}

\author{Stefano Zambito\,\orcidlink{0000-0002-4499-2545}} 
\affiliation{D\'epartement de Physique Nucl\'eaire et Corpusculaire, University of Geneva, CH-1211 Geneva 4, Switzerland}

\author{Shunliang Zhang\,\orcidlink{0009-0001-1971-8878} \PRE{\vspace*{0.1in}}} 
\affiliation{Department of Physics, Tsinghua University, Beijing, China}

\begin{abstract}
\vspace*{0.2in}
The Forward Search Experiment (FASER) at CERN's Large Hadron Collider (LHC) has recently directly detected the first collider neutrinos.  Neutrinos play an important role in all FASER analyses, either as signal or background, and it is therefore essential to understand the neutrino event rates.  In this study, we update previous simulations and present prescriptions for theoretical predictions of neutrino fluxes and cross sections, together with their associated uncertainties.  With these results, we discuss the potential for possible measurements that could be carried out in the coming years with the FASER neutrino data to be collected in LHC Run 3 and Run 4.
\end{abstract}

\maketitle 

\begin{center}
\vspace*{0.2in}
\copyright~2024 CERN for the benefit of the FASER Collaboration.  Reproduction of this article or parts of it is allowed as specified in the CC-BY-4.0 license.     
\end{center}


\clearpage

\section{Introduction} 
\label{sec:introduction}

The Forward Search Experiment (FASER)~\cite{Feng:2017uoz, FASER:2022hcn,FASER:2018bac,FASER:2020gpr} at CERN's Large Hadron Collider (LHC) complements the large LHC detectors through its ability to directly detect light, weakly-interacting particles~\cite{ FASER:2021cpr, FASER:2021ljd}.  These particles include the neutrinos of the Standard Model (SM), as well as proposed new particles.  FASER is located along the beam collision axis, 480~m from the ATLAS interaction point (IP), and began taking beam collision data at the beginning of LHC Run 3 in 2022.  

With the 2022 data set, corresponding to an integrated luminosity of $37~\ifb$, FASER detected 153 muon neutrinos, the first collider neutrinos to be directly detected~\cite{FASER:2023zcr}.  FASER also observed the first electron neutrino interactions at a collider~\cite{CERN-FASER-CONF-2023-002} and set new limits on long-lived particles~\cite{FASER:2023tle}.  The neutrinos were the most energetic neutrinos ever directly detected from an artificial source.  They have been supplemented by an additional eight muon neutrinos detected by the SND@LHC experiment~\cite{SNDLHC:2023pun}. These discoveries have opened up the new field of collider neutrino physics.  

In the coming years, FASER is expected to collect a total integrated luminosity of $250~\ifb$ in Run 3, and has recently been approved to continue operating through Run 4, which is expected to increase the total (Run 3 + Run 4) integrated luminosity to $930~\ifb$~\cite{Run4LOI}.  The large expected neutrino event rates, together with their energy and spatial distributions, will have many implications, including the potential to constrain neutrino scattering cross sections of all three flavors at unprobed energies, measure forward hadron fluxes~\cite{Kling:2023tgr}, improve constraints on parton distribution functions (PDFs) using deep-inelastic scattering (DIS) data~\cite{Cruz-Martinez:2023sdv}, resolve longstanding puzzles in astroparticle physics~\cite{Anchordoqui:2022fpn}, and test predictions for new physics~\cite{Anchordoqui:2021ghd,Feng:2022inv}.

Because neutrinos play an important role in all FASER analyses, either as the signal or as a background to new particle searches, a detailed understanding of neutrino event rates, as well as estimates of the associated uncertainties, is required. Simulations of the neutrino event rate require a number of tools and calculations that are not typically associated with colliders, since neutrino interactions have never before played a role at colliders.  In this study, the neutrino fluxes and interactions are simulated in FASER and described.  These results will be the basis for upcoming FASER analyses. 

The forward neutrino beam at the LHC mainly originates from the weak decay of the lightest mesons and baryons of a given flavor (pions, kaons, hyperons, $D$-mesons, and charm baryons). A variety of tools and calculations are available to simulate the production of these particles. The resulting neutrino flux at FASER can then be estimated using the fast neutrino flux simulation introduced in Ref.~\cite{Kling:2021gos}. That original work considered an LHC configuration (collision energy, magnet strength, and crossing angle) resembling conditions at the end of Run~2. In this work, this simulation is updated to the LHC configurations realized in Run~3 and expected in Run~4.

In addition, Ref.~\cite{Kling:2021gos} only contains a very rough estimate of the neutrino flux uncertainty. This uncertainty mainly originates from the modeling of hadron production in the primary collision, and its description was based on the event generators available at the time. Since then, additional tools and calculations of the neutrino flux have been presented. Here, these new developments are used to update neutrino flux predictions at FASER and establish the corresponding uncertainties.  

Beyond flux uncertainty, an additional source of uncertainty of the expected number of neutrino events is associated with the modeling of neutrino interactions. Although neutrino interaction cross sections in the multi-hundred GeV region have traditionally been modeled using the Bodek-Yang model~\cite{Bodek:2002vp, Bodek:2004pc, Bodek:2010km}, several new cross section models based on next-to-leading-order (NLO) structure functions have become available in recent years. These predictions will be compared, and the corresponding cross-section uncertainties will be defined.

This paper is structured as follows. A brief review of the original neutrino simulation~\cite{Kling:2021gos} and assumptions about the LHC configurations are presented in \cref{sec:simulation}. In \cref{sec:fluxes} models for forward hadron production and their implications for neutrino fluxes are compared, and in \cref{sec:interactions} neutrino interactions and the accompanying uncertainties are discussed. With the provided neutrino flux and cross-section results, predictions for neutrino event rates at FASER are presented in \cref{sec:distributions}. This includes energy and spatial distributions for all three neutrino flavors. Additionally, potential measurements that could be conducted with FASER in LHC Run 3 and Run 4 are briefly discussed. The conclusions derived from this analysis are summarized in \cref{sec:conclusion}. In the Appendix several forward charm production models are compared to each other and to data.

\section{Simulation of Forward Neutrinos at the LHC} 
\label{sec:simulation}

The beam of forward, high-energy neutrinos observable at FASER mainly originates from the weak decays of hadrons that are produced at the ATLAS interaction point.\footnote{Neutrinos can also be produced in downstream hadronic showers resulting from collisions of primary hadrons with the LHC infrastructure. However, given the typically lower energy and large spread of hadrons in later stages of the shower, as well as the fact that these hadrons are more likely to interact in the LHC infrastructure than to decay to neutrinos, the resulting neutrino flux is subdominant. Indeed, as found in Ref.~\cite{Kling:2021gos}, the contribution of such processes to the neutrino flux is below the percent level at $\sim \tev$ energies for all flavors.  Based on this finding, this flux component is not considered in our neutrino flux estimate here.} This includes light hadrons (pions, kaons, and hyperons), which are long-lived and decay inside the LHC vacuum beam pipe, and also charm hadrons, which decay essentially promptly. 

To obtain the neutrino flux, one needs to model the trajectory of the long-lived hadrons through the LHC beam pipe and magnetic fields and also model the decay of these hadrons into neutrinos. This is done using the fast neutrino flux simulation introduced in Ref.~\cite{Kling:2021gos}. This fast neutrino flux simulation (i) reads the forward hadron fluxes from \texttt{HepMC} files produced by the Monte-Carlo (MC) event generator; (ii) propagates the long-lived hadrons through the LHC beam pipe and magnetic fields; (iii) obtains the neutrinos from decays of hadrons at multiple locations along their trajectory; and (iv) stores the resulting neutrinos going through a sample plane at the FASER location as a MC event sample. All parts of the outlined simulation are implemented as a \texttt{RIVET} module~\cite{Buckley:2010ar, Bierlich:2019rhm}. The results of the fast simulation have been validated against the full simulation using \texttt{BDSIM},\footnote{\texttt{BDSIM}~\cite{Nevay:2018zhp} 
is a code based on \texttt{Geant4}~\cite{Agostinelli:2002hh}, \texttt{ROOT}, and \texttt{CLHEP} to create radiation transport models of accelerators that can track all particles. It creates a \texttt{Geant4} model with translation to a curvilinear coordinate system that follows the accelerator, as well as more accurate and faster tracking algorithms specific to the magnetic fields of an accelerator. Custom component geometry can be combined with a library of detailed LHC and generic magnet geometries to create complete accelerator models tracking all particles using the full physics of \texttt{Geant4}, including in the yokes of magnets. \texttt{BDSIM} tracking has be thoroughly validated and it is used extensively in the accelerator community and at CERN.} and the predictions were found to be in good agreement~\cite{Kling:2021gos}. BDSIM accounts for all contributions to the neutrino flux, including those arising from hadronic interactions in the beam pipe and surrounding material; see Sec.~2.8 of Ref.~\cite{Feng:2022inv} for a description of BDSIM in the context of forward physics. In particular, the differences between the full and fast simulations are significantly smaller than the differences between MC event generators for neutrino energies above several hundred GeV, which are the focus of both SM and beyond the SM (BSM) studies so far.\footnote{In the future, if neutrino events with lower energies become important, further study of the discrepancies between the fast simulation, BDSIM, and other transport codes, such as FLUKA, are warranted.}

The simulation of Ref.~\cite{Kling:2021gos} assumed the LHC Run~2 configuration with center-of-mass energy $\sqrt{s} = 13~\tev$. For this work, the \texttt{RIVET} module has been updated for center-of-mass energies of 13.6 TeV and 14 TeV for Run 3 and Run 4, respectively.  In particular, the strengths of the magnetic fields were adjusted to the higher beam energies. In addition, the planned major changes to the LHC infrastructure for Run 4~\cite{ZurbanoFernandez:2020cco} (including the geometry of the beam pipe, magnet configuration, and position of the target neutral absorber) were incorporated, as illustrated in \cref{fig:geometry}. The magnet strengths as well as the aperture information are taken from the \texttt{BDSIM} model, which is automatically prepared from the MAD-X~\cite{MAD-X} optics strengths in combination with a separate aperture model and detailed models of many components. 

The simulation of forward neutrinos also depends critically on the beam crossing angle, which modifies the nominal line-of-sight (LOS), that is, the LOS in the absence of a crossing angle, to the true, or actual, LOS, that is, the LOS with the crossing angle included.  The previous simulation used the 2018 Run 2 beam crossing half-angle of $\theta_{1/2} = 150~\mu$rad vertically upwards.  For Run 3, the beam crossing half-angle at the ATLAS IP was 160~$\mu$rad downwards in 2022 and 2023, and it is expected to change to 160~$\mu$rad upwards for 2024 and 160~$\mu$rad horizontally for 2025.\footnote{The crossing angle is also changed by a few 10 $\mu$rad during each physics fill, but this has a negligible effect on the results for FASER and is not taken into account in this work.} For simplicity, the estimates labeled Run 3 in this paper assume the 2022 and 2023 crossing half-angle $\theta_{1/2} = 160~\mu$rad vertically downwards. The changes expected for 2024 and 2025 do not significantly modify the estimates of the number of neutrino interactions in FASER. More details can be found in Ref.~\cite{Run4LOI}.  This Run 3 crossing angle shifts the true LOS downward by 7.7 cm from the nominal LOS.  For Run 4, we assume $\theta_{1/2} = 250~\mu$rad~\cite{TomasGarcia:2803611} in the horizontal plane (away from the LHC ring), which shifts the true LOS 12 cm horizontally from the nominal LOS. 

\begin{figure}[tbp]
\includegraphics[width=0.49\textwidth]{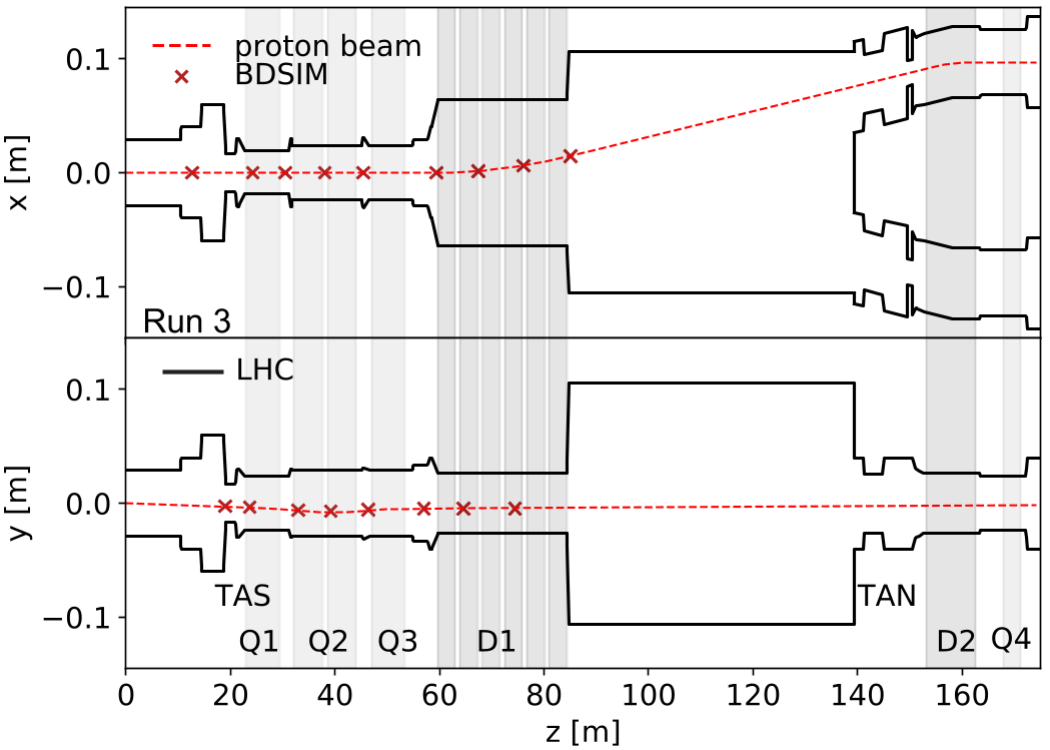} 
\includegraphics[width=0.49\textwidth]{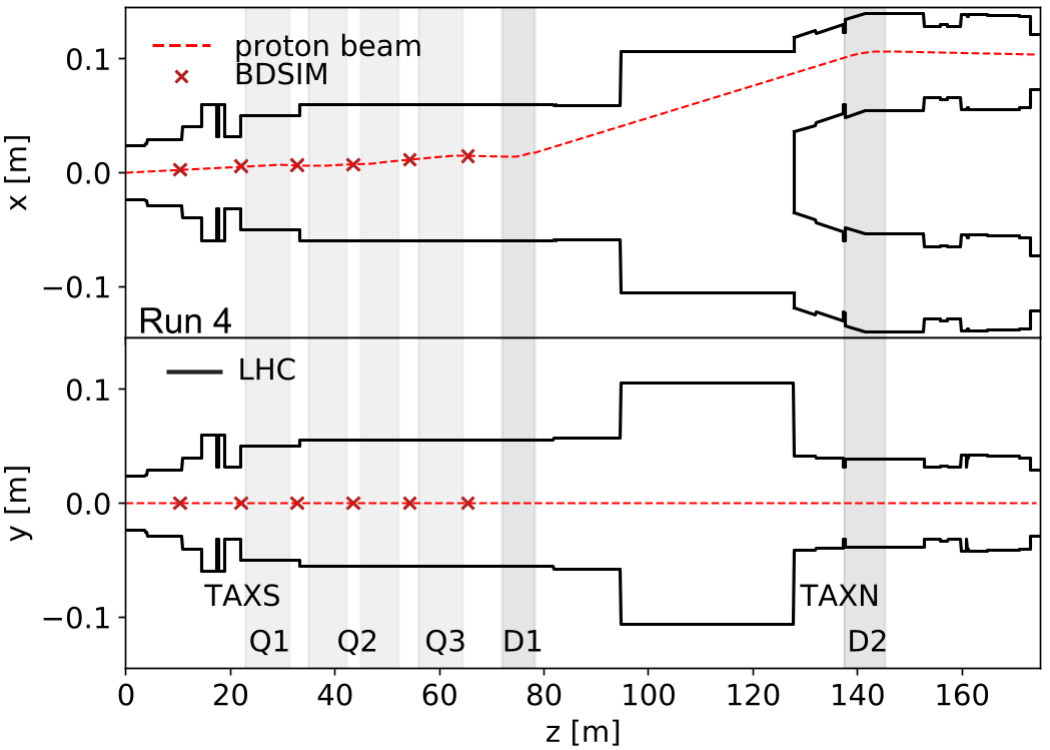} 
\caption{Beam Pipe Geometry and Magnets. The boundaries of the LHC’s beam pipe (black lines) and magnetic fields (gray shaded areas) assumed in the geometric model for LHC Run~3 (left) and Run~4 (right). The $x$, $y$, and $z$ coordinates form a right-handed FASER coordinate system (see text) with the ATLAS IP at the origin, and FASER at approximately $(0, 0, 480~\m)$.  The red lines show trajectories calculated by the updated simulation from Ref.~\cite{Kling:2021gos} of the outward-going 6.8 TeV proton beam at Run~3 (left) and the 7 TeV proton beam at Run~4 (right). The markers show the tracking points obtained using \texttt{BDSIM}. The TAN and TAS (TAXN and TAXS) are the LHC (high luminosity LHC) target neutral and passive absorbers.
\label{fig:geometry}
}
\end{figure}

The FASER$\nu$ detector consists of 730 1.1-mm thick tungsten plates interleaved with emulsion films, with a total target mass of 1.1 tonnes, and transverse dimensions 25 cm wide and 30 cm high.  In our calculations, FASER$\nu$ is simulated by assuming a simplified detector that is 25 cm wide, 30 cm high, and 80 cm deep in the beam direction, and filled with 1.1 tonnes of tungsten.  

To specify the location of the detector, FASER uses a right-handed coordinate system, with the positive $x$-axis pointing in the horizontal direction toward the center of the LHC, the positive $y$-axis pointing vertically upward, and the positive $z$-axis pointing from the ATLAS IP toward FASER. The nominal LOS is at the origin $(x, y) = (0,0)$.  The simplified detector's location in the transverse plane is matched to FASER$\nu$'s location during 2022/2023~\cite{FASER:2023zcr}, which is centered at $(x, y) = (1.0~\cm, -3.3~\cm)$. Given this location and the 2022/2023 beam crossing angle, FASER$\nu$ covers pseudorapidities $\eta > 8.3$.  Note that the center of the FASER spectrometer, defined by the axis of symmetry of the magnets, is at $(0, -1.2~\cm)$.  For Run 4, the FASER and FASER$\nu$ detectors are both assumed to be shifted 5.0 cm horizontally away from the LHC relative to their 2022/2023 locations, which brings the centers of these detectors closer to the true LOS~\cite{Run4LOI}.  For Run 4, then, the center of FASER$\nu$ is at $(6.0~\cm, -3.3~\cm)$, and the center of FASER is at $(x,y) = (5.0~\cm, -1.2~\cm)$.  With the beam crossing angle discussed above, FASER$\nu$ covers pseudorapidities $\eta > 8.2$ in Run 4.

\section{Neutrino Fluxes and Uncertainties}
\label{sec:fluxes}

As already discussed, the two primary components of the forward LHC neutrino beam observable at FASER are neutrinos produced downstream in light hadron decays and neutrinos produced promptly in charm hadron decays. Aside from differences in the production location and the associated simulation requirements, they also differ qualitatively in their theoretical modeling.  We, therefore, discuss them in turn.

In inelastic collisions at the LHC, forward light hadrons are commonly produced. Most of these collisions are of a soft, low-scale nature, with a characteristic energy scale $Q$ roughly equivalent to $\Lambda_{\text{QCD}}$, and these collisions generally don't result in the production of heavy or large-transverse-momentum particles. The kinematics of these events falls outside the scope of perturbative QCD's applicability. As a result, these events are often simulated using phenomenological hadronic interaction models. These models vary significantly in several aspects, such as their underlying theoretical framework and the methods they use to represent hadronization, parton distributions, diffraction, and correlations. (For an overview, see Table 2 of Ref.~\cite{Albrecht:2021cxw}.)

Several such tools have been developed for cosmic ray physics. The most up-to-date event generators include \texttt{EPOS-LHC}~\cite{Pierog:2013ria}, \texttt{SIBYLL~2.3d}~\cite{Riehn:2019jet}, and \texttt{QGSJET~2.04}~\cite{Ostapchenko:2010vb}; these version numbers are implicit when omitted below.  In addition, a new tune of \texttt{Pythia~8.3}~\cite{Sjostrand:2014zea,Bierlich:2022pfr} has recently been presented that is specifically designed to describe forward particle production at the LHC~\cite{Fieg:2023kld}; this tune will be referred to as \pythiaF. 

All of the mentioned event generators have been tuned to or validated with a variety of low- and high-energy accelerator data. These include measurements of forward neutral hadron production at LHCf~\cite{LHCf:2008lfy}, a zero-degree calorimeter with two detectors that are located about 140~m upstream and downstream of ATLAS, covering pseudorapidities $|\eta| \agt 8.8$. In \cref{fig:LHCf-photon}, the predictions of the event generators are compared to the forward photon~\cite{LHCf:2017fnw}, $\eta$-meson~\cite{Piparo:2023yam}, and neutron~\cite{LHCf:2018gbv} energy spectra measured by LHCf at $\sqrt{s} = 13~\tev$. Here the photons originate primarily from neutral pion decay. No production model gives a perfect fit to the data for all particles, energies, and pseudorapidities, but the four event generators shown all provide fairly good descriptions of the data, and together they form an envelope around most of the data. Notable exceptions include the very forward photon spectrum at $\eta>10.94$ and the very forward neutron spectrum. However, the photon spectrum in this region corresponds to a small radius of $r < 1~\text{cm}$ at FASER, and there is also a conflicting pion measurement at $\sqrt{s} = 7~\tev$~\cite{LHCf:2015rcj}, where the generators overestimate the pion flux. On the other hand, neutrons only contribute to the neutrino flux through secondary interactions, which are negligible for $\mathcal{O}(\tev)$ neutrino energies.

\begin{figure}[tbp]
\includegraphics[width=0.32\textwidth]{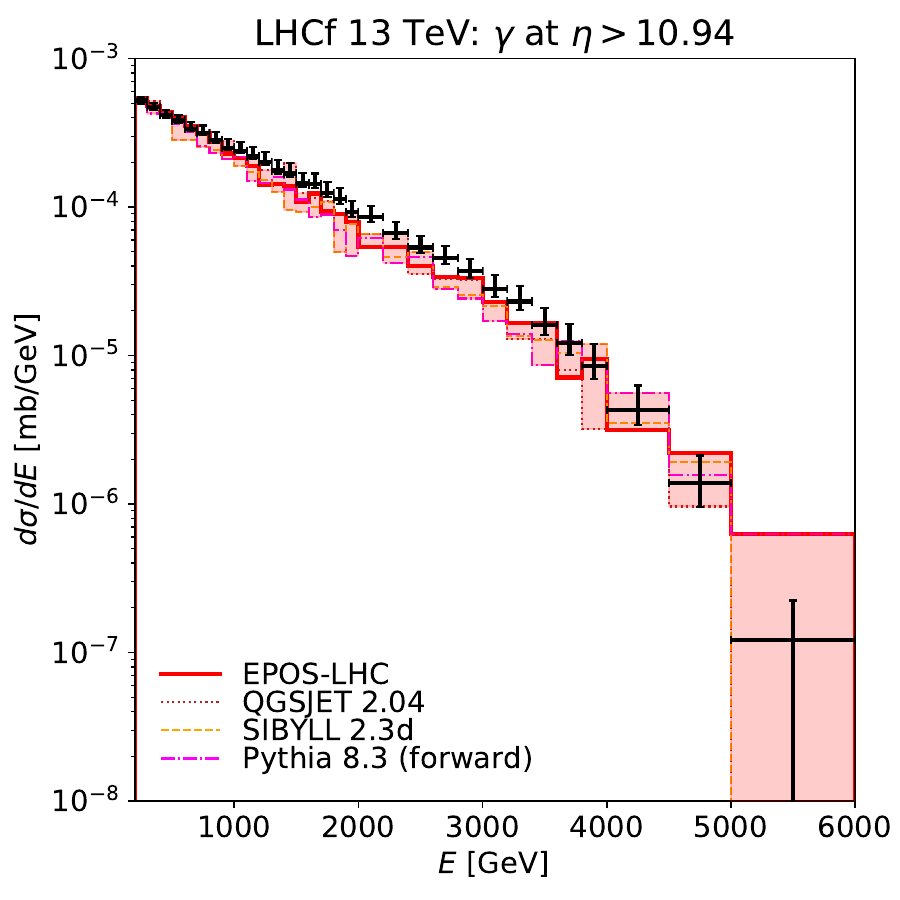} 
\includegraphics[width=0.32\textwidth]{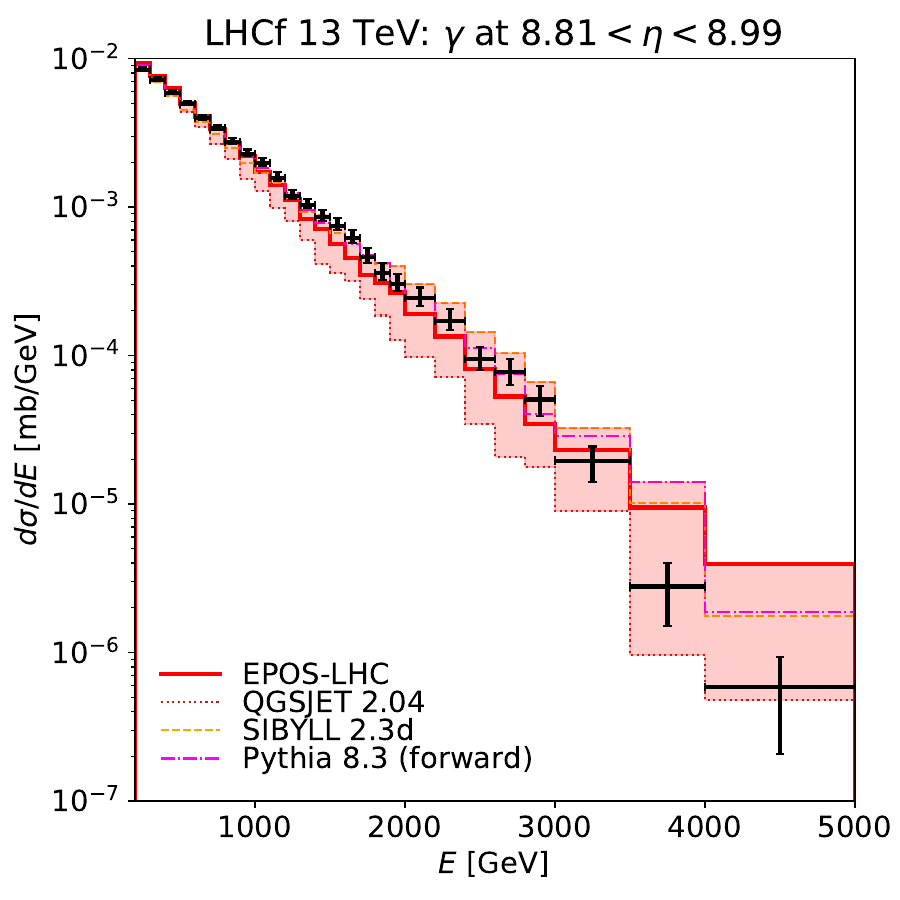} 
\includegraphics[width=0.32\textwidth]{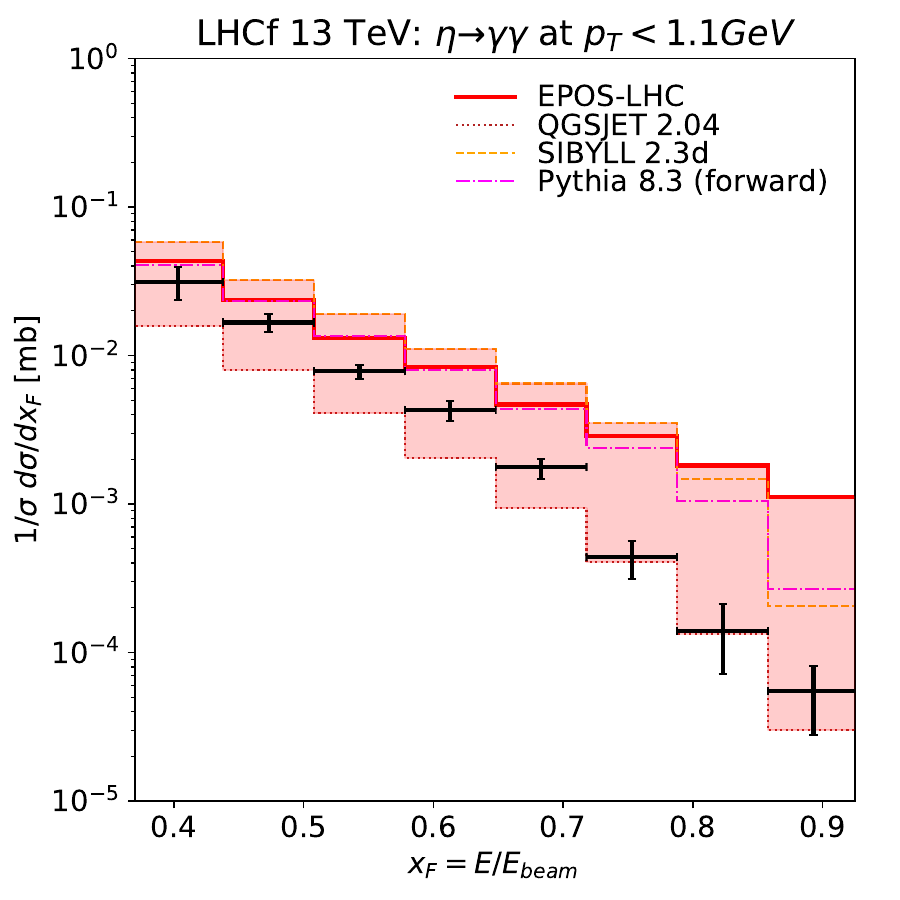} 
\includegraphics[width=0.32\textwidth]{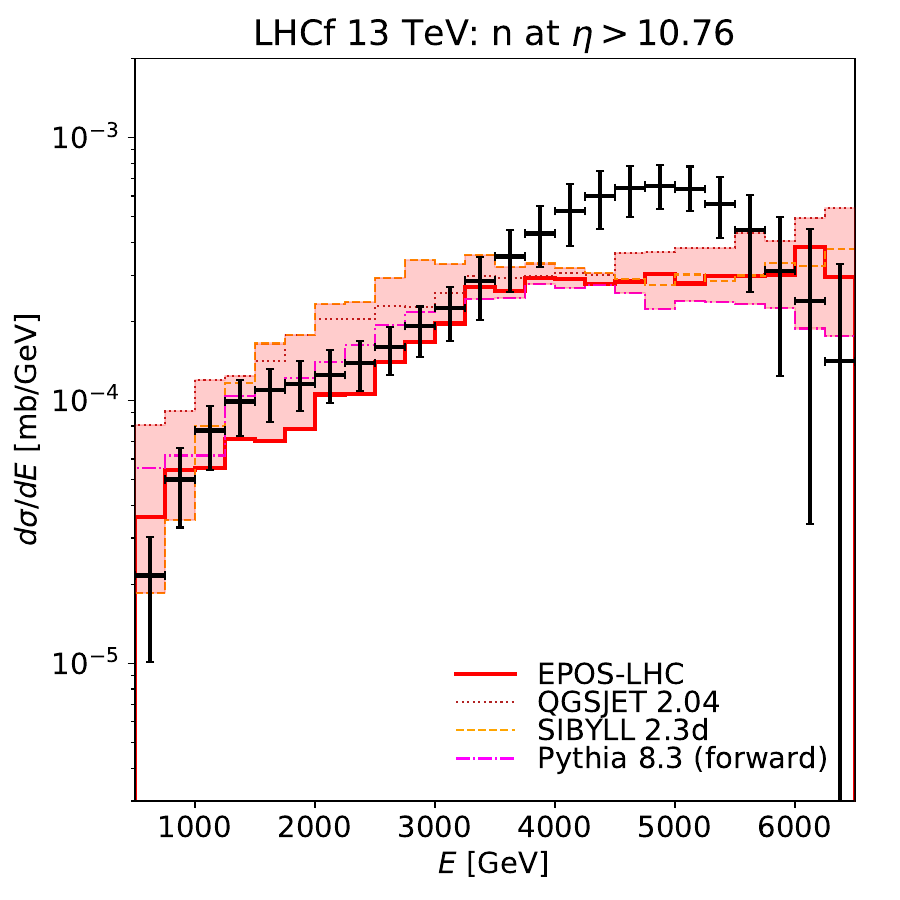} 
\includegraphics[width=0.32\textwidth]{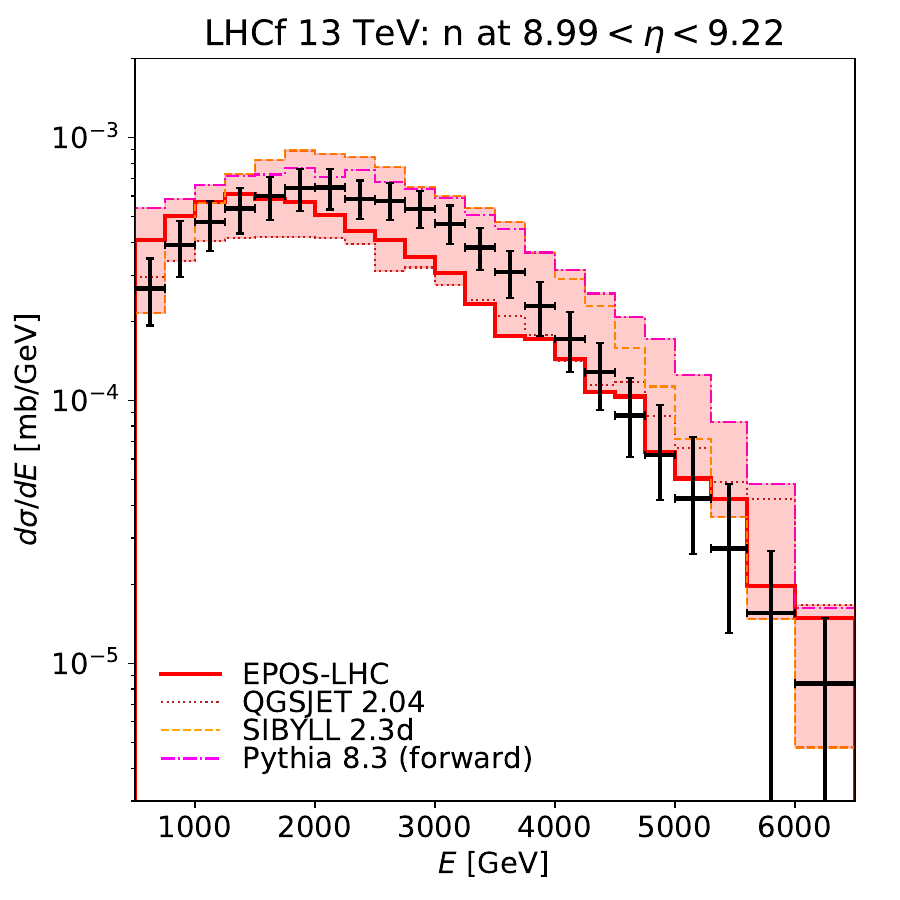} 
\includegraphics[width=0.32\textwidth]{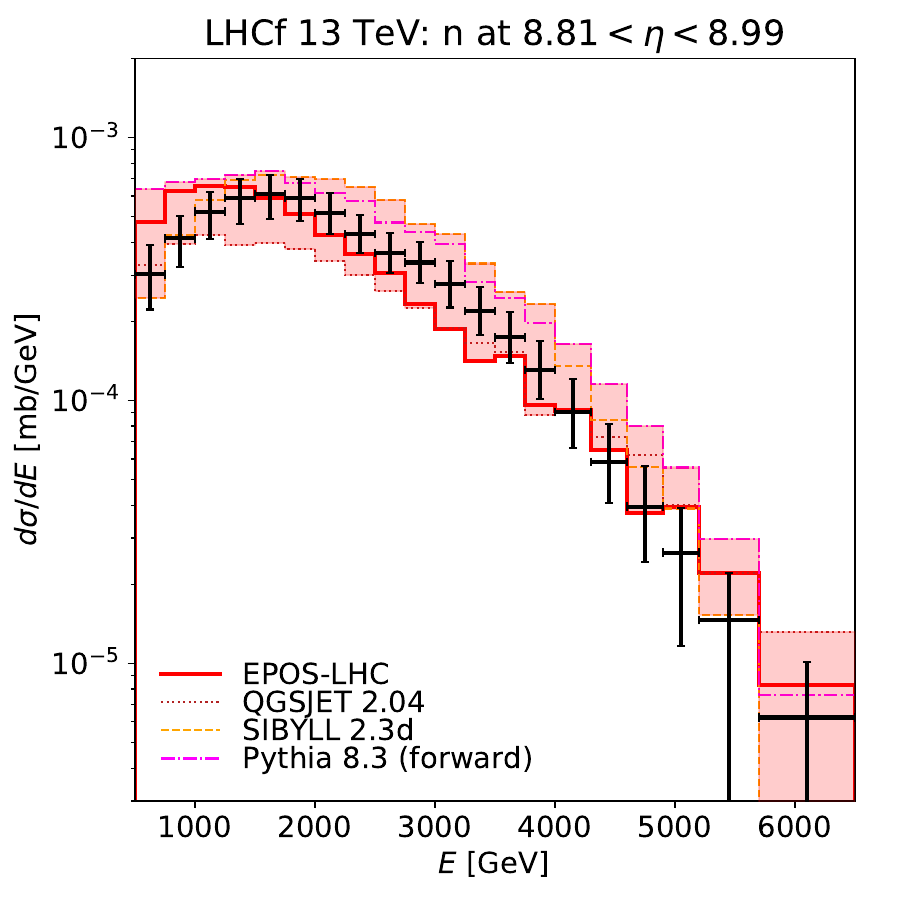} 
\caption{Forward Particle Energy Spectra at LHCf and Model Predictions. The energy spectra of forward photons~\cite{LHCf:2017fnw} (upper left and upper center), $\eta$-mesons~\cite{Piparo:2023yam} (upper right), and neutrons~\cite{LHCf:2018gbv} (lower left, lower center, and lower right) measured by LHCf in different pseudorapidity bins, compared to the predictions of the event generators \texttt{EPOS-LHC}~\cite{Pierog:2013ria}, \texttt{SIBYLL}~\cite{Riehn:2019jet}, \texttt{QGSJET}~\cite{Ostapchenko:2010vb}, and \pythiaF~\cite{Fieg:2023kld}. The shaded bands correspond to the spread of the event generator predictions. 
\label{fig:LHCf-photon}
}
\end{figure}

Most of the event generators only provide a central prediction, with no measure of uncertainty. To define an associated uncertainty, we follow an approach that is often adopted in astroparticle physics, that is, the spread of event generator predictions is taken as an estimator of the production uncertainty. In particular, we consider the spread in the four aforementioned event generators, which have the best agreement with LHCf data: \texttt{EPOS-LHC}, \texttt{SIBYLL}, \texttt{QGSJET}, and \pythiaF. This approach has the advantage that it captures differences associated with the underlying physics modeling. It should be noted that an alternative definition of uncertainties, using tuning variations in \texttt{Pythia~8.3}, has been proposed in Ref.~\cite{Fieg:2023kld}. That study found that the uncertainties obtained in this way are similar to those obtained using the spread of event generators.

In \cref{fig:nu_flux}, the combined energy spectra of charged-current(CC)-interacting electron and muon neutrinos, summed with their corresponding anti-neutrinos, that are produced in light hadron decays and interact in FASER$\nu$ in LHC Run 3 are shown in red.  Following the above discussion, \texttt{EPOS-LHC} is used as the central prediction and the envelope formed by \texttt{EPOS-LHC}, \texttt{SIBYLL}, \texttt{QGSJET}, and \pythiaF is used to define an uncertainty band. These results depend on the assumed interaction cross section. As will be discussed in detail in \cref{sec:interactions}, we use the neutrino interaction generator \texttt{GENIE 3.4}~\cite{GENIE:2021wox} to determine the total neutrino interaction cross section. 

\begin{figure}[tbp]
\includegraphics[width=0.49\textwidth]{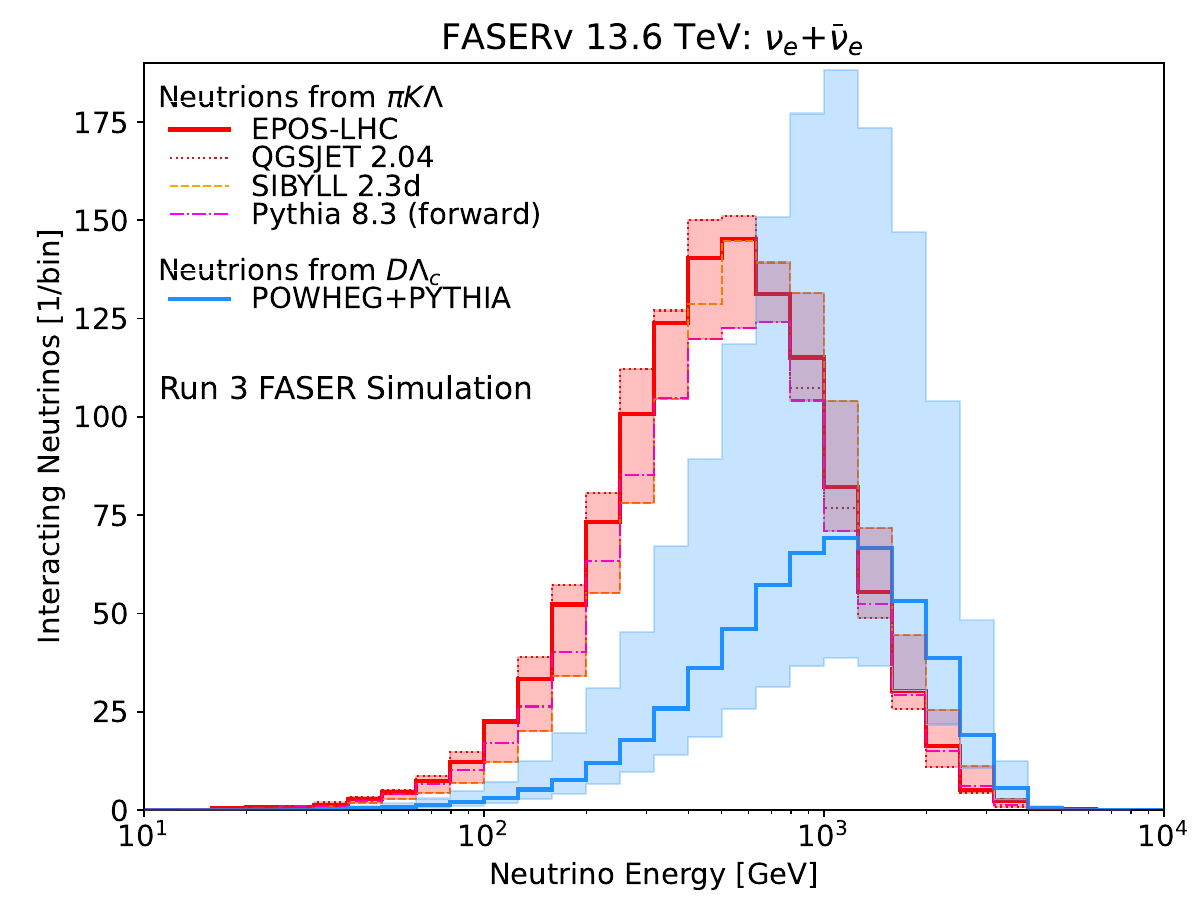}
\includegraphics[width=0.49\textwidth]{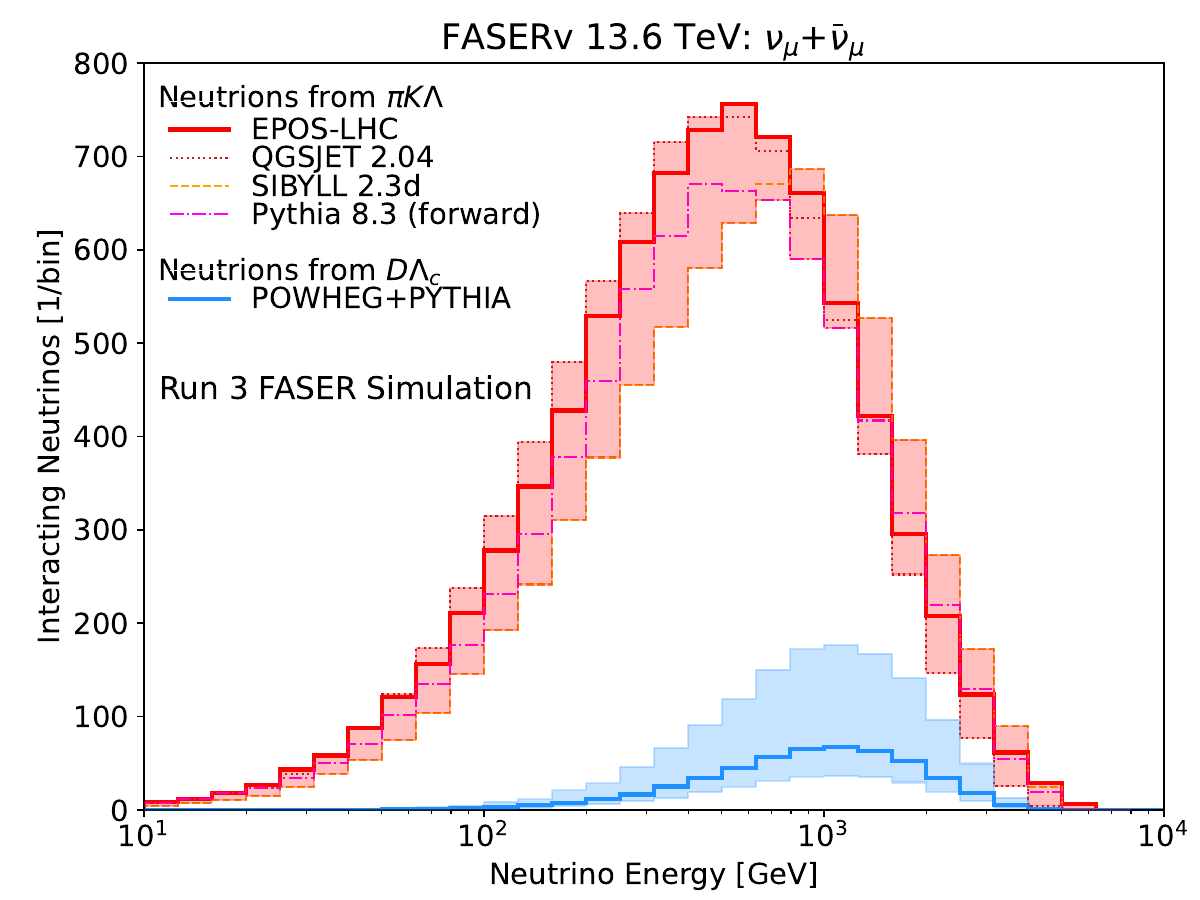}
\caption{Neutrino CC Interactions in FASER$\nu$. The energy spectrum of electron neutrinos (left) and muon neutrinos (right) expected to have CC interactions in FASER$\nu$ in LHC Run 3 with an integrated luminosity of $250~\ifb$.  The component of neutrinos originating from light (charm) hadron decays is shown in red (blue).  The solid contours are the central values, and the shaded regions show the corresponding uncertainties (see text).  
\label{fig:nu_flux}
}
\end{figure}

The situation is different for forward charm hadrons. Their production is, so far, only included in some of the available hadronic interaction models. Ref.~\cite{Kling:2021gos} used \texttt{SIBYLL 2.3d}~\cite{Riehn:2019jet}, \texttt{Pythia~8.2}~\cite{Sjostrand:2014zea}, and \texttt{DPMJET~3.2019.1}~\cite{DPMJET} (an update of Ref.~\cite{Roesler:2000he}) and found that their predictions may differ by large factors of ${\cal O}(10)$. In contrast to light mesons, forward charm production can, in principle, be described using perturbative QCD methods. Although several such predictions based on analytic perturbative calculations have been presented~\cite{Bai:2020ukz, Maciula:2022lzk, Bhattacharya:2023zei}, these often use approximate descriptions of either the hard scattering or the hadronization that may affect their reliability.

More recently, a new calculation was presented that simulates forward charm production in a way that addresses the shortcomings of previous estimates. It uses state-of-the-art QCD predictions for heavy hadron production that include radiative corrections~\cite{Buonocore:2023kna}, as well as the sophisticated modeling of hadronization implemented in MC generators. This simulation uses \texttt{POWHEG}~\cite{Nason:2004rx,Frixione:2007vw,Alioli:2010xd} with the NNPDF3.1sx+LHCb PDF set~\cite{Ball:2017otu, Bertone:2018dse} to model charm production at next-to-leading order in $\alpha_s$ with small-$x$ resummation at next-to-leading logarithmic accuracy, and then matches it with \texttt{Pythia~8.3}~\cite{Bierlich:2022pfr} for parton showering and hadronization. This analysis also includes an uncertainty estimate based on scale uncertainties. Additional sources of uncertainties, for example, those arising from the modeling of hadronization, were investigated, but were found to be within the scale uncertainty band. We therefore use the scale variations to define the flux uncertainties. 

In \cref{fig:nu_flux}, the predicted energy spectra of electron and muon neutrinos that are produced in charm hadron decays and interact in FASER$\nu$ in LHC Run 3 are shown in blue.  Following the above discussion, \texttt{POWHEG + Pythia 8.3} with default scales is used as the central prediction, and the resummation and factorization scales are varied by a factor of two to define the uncertainty band.

\section{Neutrino Interaction Cross Sections}
\label{sec:interactions}

The number of neutrino events depends not only on the neutrino flux, but also the neutrino interaction cross section.  Given the typically large neutrino energy $E_\nu >100~\gev$, most neutrino interactions at FASER can be described as DIS. It is worth noting, however, that, especially at lower energies $E_\nu<100~\gev$, there can be a substantial non-DIS contribution. Following the notation of Ref.~\cite{Candido:2023utz}, the CC neutrino-nucleon interaction cross section can be written as
\be
\frac{d\sigma_{\nu N}}{dx \, dy} = \frac{G_F^2 m_N E_\nu}  {\pi(1+m_{W}^2/Q^2)^2} \, \big[ xy^2 F_1  + (1-y) F_2  +  xy(1-y/2) F_3] \ , 
\ee
where $x$ is the fraction of the nucleon's momentum carried by the quark in the initial state, $y$ is the fraction of the neutrino momentum transferred to the hadronic system, $m_N$ and $m_W$ are the masses of the nucleon and $W$ boson, respectively, $Q^2= 2 E_\nu m_N x y$ is the transferred four-momentum, and $F_{i}(x,Q^2)$ are the structure functions of the proton. Different models for the structure functions have been proposed in the literature, some of which allow one to extend this formalism into the non-DIS regime. 

\begin{figure}[tbp]
\includegraphics[width=0.49\textwidth]{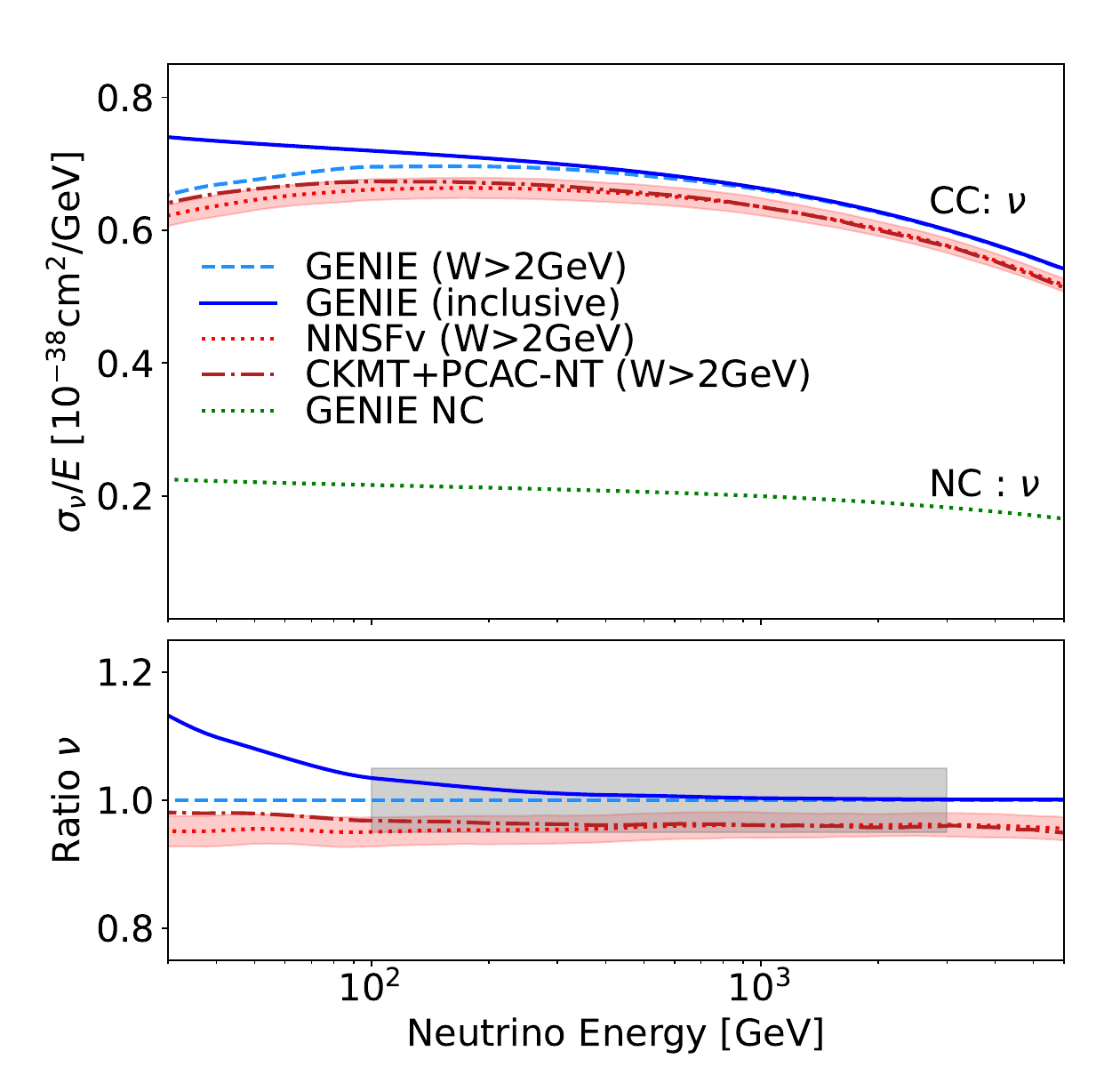}
\includegraphics[width=0.49\textwidth]{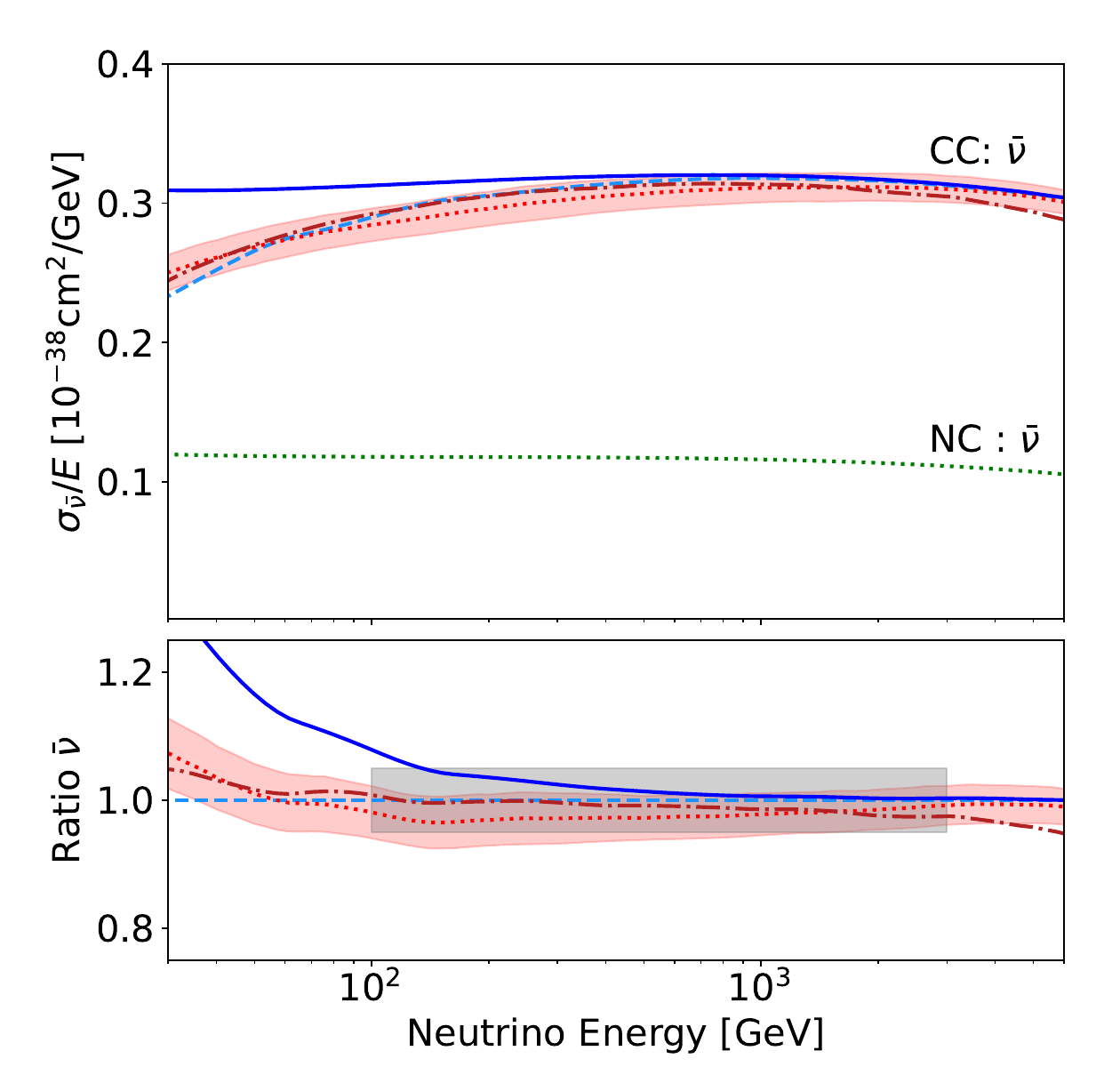}
\caption{Top: Muon neutrino (left) and anti-neutrino (right) CC interaction cross sections as functions of the incoming neutrino energy, as obtained using \texttt{GENIE}~\cite{GENIE:2021wox}, \texttt{NNSFv}~\cite{Candido:2023utz}, and \texttt{CKMT+PCAC-NT}~\cite{Jeong:2023hwe}, as indicated. For comparison, we show the neutral-current (NC) cross section prediction of \texttt{GENIE} in dotted green. The electron neutrino cross section prediction is approximately the same, differing by less than 0.1\% for $E_{\nu}>100~\gev$, and the tau neutrino cross section is $\lesssim 20\%$ smaller, due to the tau mass. The red-shaded bands show the \texttt{NNSFv} uncertainty band. Bottom: Muon neutrino (left) and anti-neutrino (right) CC cross sections normalized to the \texttt{GENIE} $W > 2~\gev$ results.  The gray bands correspond to flat 6\% uncertainties in the energy range $100~\gev < E_\nu < 3~\tev$. }
\label{fig:crosssections}
\end{figure}

The Bodek-Yang model~\cite{Bodek:2002vp, Bodek:2004pc, Bodek:2010km} is widely recognized as a phenomenological framework for describing inelastic neutrino-nucleon scattering cross sections in the multi-GeV energy range. It is implemented in \texttt{GENIE}~\cite{GENIE:2021wox}, a commonly used neutrino interaction MC generator, which has been extensively benchmarked in neutrino experiments in the 0.1--100 GeV energy range and simulates both DIS and non-DIS contributions to the cross section.  The Bodek-Yang model builds upon structure functions using effective leading-order GRV98 PDFs~\cite{Gluck:1998xa}.  To account for mass, higher-order QCD, and nuclear effects, various phenomenological corrections are employed. In addition, modifications have been incorporated to extend its applicability to the non-perturbative, low-$Q^2$ regime. 

For FASER predictions, however, the Bodek-Yang model has certain limitations: (i) it relies on obsolete PDFs that neglect recent constraints on proton and nuclear structure obtained in the last 25 years; (ii) it only includes PDFs for up, down, and strange quarks; (iii) it omits available higher-order QCD calculations; (iv) it was primarily designed for the multi-GeV domain rather than TeV neutrino energies.

A variety of other cross section calculations based
on NLO structure functions and modern PDF sets have been presented, primarily for applications in astroparticle physics, including the CSMS~\cite{Cooper-Sarkar:2011jtt} and BGR18~\cite{Bertone:2018dse, Garcia:2020jwr} models, as well as cross sections based on the CT18~\cite{Xie:2023suk} PDFs.  More recently, additional models have been presented that further extend the NLO structure function models to the low $Q^2$ regime and hence extend their applicability to lower neutrino energies. Examples of such models are \texttt{NNSFv}~\cite{Candido:2023utz} and \texttt{CKMT+PCAC-NT}~\cite{Jeong:2023hwe}, which have good agreement with CSMS for large $E_{\nu}$ when DIS becomes dominant. For \texttt{NNSFv}, the structure functions are determined by a data-driven parametrization at low and moderate values of $Q^2$ matched to perturbative QCD calculations at high $Q^2$. Notably, \texttt{NNSFv} also includes an uncertainty estimate obtained in a data-driven way.  \texttt{CKMT+PCAC-NT} uses a structure function parametrization augmented by a correction to account for the partial conservation of the axial-vector current, normalized to structure functions evaluated at NLO order in QCD, and includes target mass and heavy quark corrections.  We note, however, that the NLO structure function models do not describe neutrino scattering at low-$W$, where $W$ is the invariant mass of the final-state hadronic system, and are not yet available in event generators. We can, however, use them to validate the Bodek-Yang model implemented in \texttt{GENIE}.

In \cref{fig:crosssections}, the neutrino CC interaction cross sections obtained by the different approaches are compared, including the inclusive interaction cross sections obtained using \texttt{GENIE}, and the predictions of \texttt{GENIE}, \texttt{NNSFv}, and \texttt{CKMT+PCAC-NT} after the DIS selection cuts $Q > 0.03~\gev$ and $W>2~\gev$.  The red-shaded band shows the \texttt{NNSFv} uncertainty, as obtained in Ref.~\cite{Candido:2023utz}. There is general agreement between the predictions of the Bodek-Yang model, as implemented in \texttt{GENIE}, with \texttt{NNSFv} and \texttt{CKMT+PCAC-NT}, and the cross section uncertainties for neutrinos with energies above $100~\gev$ are roughly at the 6\% level. For our calculations of the interacting neutrino rate, the default GENIE cross section is used. For purposes of comparison, the NC cross sections predicted by \texttt{GENIE} are also shown in \cref{fig:crosssections}.

\section{Neutrino Rates and Distributions}
\label{sec:distributions}

We now turn to our predictions for the forward neutrino spectra at the LHC during Run 3 and Run 4. The hadron spectra are generated using the event generators discussed in \cref{sec:fluxes}, propagated down the beam pipe, and decayed to produce a flux of neutrinos. The total neutrino cross section provided by the \texttt{GENIE} implementation of the Bodek-Yang model, as discussed in \cref{sec:interactions}, is used to produce the energy spectra of CC neutrino interactions in FASER$\nu$.

\renewcommand{\arraystretch}{1.5}
\setlength{\tabcolsep}{5pt}
\begin{table*}[bp]
\centering
\begin{tabular}{c|c||c|c|c||c|c|c}
  \hline
  \hline
  \multicolumn{2}{c||}{Generators} & 
  \multicolumn{3}{c||}{FASER$\nu$ at Run~3} & 
  \multicolumn{3}{c}{FASER$\nu$ at Run~4} \\
  \hline
  light hadrons & charm hadrons
  & $\nu_e+\bar\nu_e$    
  & $\nu_\mu+\bar\nu_\mu$    
  & $\nu_\tau+\bar\nu_\tau$    
  & $\nu_e+\bar\nu_e$    
  & $\nu_\mu+\bar\nu_\mu$    
  & $\nu_\tau+\bar\nu_\tau$    
  \\
  \hline
  \hline
    \texttt{EPOS-LHC} & --
  & 1149
  & 7996
  & --
  & 3382 
  & 23054
  &  --
  \\
    \texttt{SIBYLL~2.3d}  & --
  & 1126
  & 7261
  & --    
  & 3404
  & 21532
  & --
  \\
    \texttt{QGSJET~2.04} & -- 
  & 1181
  & 8126
  &  --
  & 3379
  & 22501
  & --
  \\
    \pythiaF & -- 
  & 1008
  & 7418
  & --
  & 2925
  & 20508
  & --
  \\
  \hline
    -- & \texttt{POWHEG} 
    Max
  & 1405
  & 1373
  &  76
  & 4264
  & 4068
  & 255
  \\
    -- & \texttt{POWHEG} 
  &  527
  &  511
  & 28
  &  1537
  & 1499
  & 91
  \\
    -- & \texttt{POWHEG} 
    Min
  & 294
  &  284
  & 16
  &  853
  & 826
  &  51
  \\
  \hline  
  \hline
  \multicolumn{2}{c||}{Combination}
  & $1675_{-372}^{+911}$
  & $8507_{-962}^{+992}$
  & $28_{-12}^{+48}$
  & $4919_{-1141}^{+2748}$
  & $24553_{-3219}^{+2568}$
  & $91_{-41}^{+163}$
  \\
  \hline
  \hline
\end{tabular}
\caption{The expected number of CC neutrino interaction events occurring in FASER$\nu$ during LHC Run~3 with $250~\ifb$ and Run~4 with $680~\ifb$. The detector geometry and locations for Run 3 and Run 4 are as described in \cref{sec:simulation}, and results are shown for the various event generators described in \cref{sec:fluxes}. In the bottom row, for the combination, we show the sum of the averages of the light hadron and charm hadron contributions as the central prediction, and their spread as the uncertainty.}
\label{tab:interactions}
\end{table*}

In \cref{tab:interactions} the total number of neutrinos interacting in FASER$\nu$ is shown for each flavor in LHC Run 3 and Run 4.  For neutrinos produced in light hadron decay, results for \texttt{EPOS-LHC}, \texttt{SIBYLL}, \texttt{QGSJET}, and \pythiaF are displayed.  The results from these event generators agree within roughly 10\%.  For neutrinos produced in charm hadron decay, results are shown for \texttt{POWHEG+Pythia 8.3} and the scale variations discussed in Ref.~\cite{Buonocore:2023kna}, providing a maximum, central, and minimum prediction for the charm hadron flux.  The spread in event rates is much larger for charm hadrons than for light hadrons, as seen in the lower section of \cref{tab:interactions}. The charm hadrons are the source of approximately 30\% of the $\nu_e$ event rate, 5\% of the $\nu_{\mu}$ event rate, and 100\% of the $\nu_\tau$ event rate. Moreover, the fraction of $\nu_e$ coming from charm hadron decay is large at higher neutrino energies and is approximately 50\% at $E_{\nu}=1~\tev$ and 90\% at $E_{\nu}=3~\tev$.  In the bottom row, the central prediction is derived by summing the \texttt{EPOS-LHC} contribution for light hadrons and \texttt{POWHEG+Pythia 8.3} for charm hadrons, while their variation is used to estimate the uncertainty. Overall, the $\nu_e$, $\nu_{\mu}$, and $\nu_{\tau}$ event rates are found to be approximately 1700, 8500, and 30 in LHC Run 3 and 4900, 25000, and 90 in LHC Run 4, respectively, with the uncertainty in each being dominated by the uncertainty in charm hadron production. 

In the upper part of \cref{fig:fasernu_spectra}, the interacting neutrino spectra is shown, including uncertainties, for all flavors at FASER$\nu$ for an integrated luminosity of $250~\ifb$ for the Run 3 configuration. The central prediction is obtained using \texttt{EPOS-LHC} for neutrinos from light hadrons and \texttt{POWHEG+Pythia 8.3} for charm hadrons.  For light hadron production, the uncertainty is defined as the spread of event generators \texttt{EPOS-LHC}, \texttt{SIBYLL}, \texttt{QGSJET}, and \pythiaF. For charm hadrons, the error bands are obtained using scale uncertainties from \texttt{POWHEG+Pythia~8.3}.  The central panel shows the same spectra, but normalized to the central prediction. The lower panel shows the fraction of neutrinos produced in charm hadron decays.

\begin{figure}[tbp]
\centering
\includegraphics[width = 0.46\textwidth]{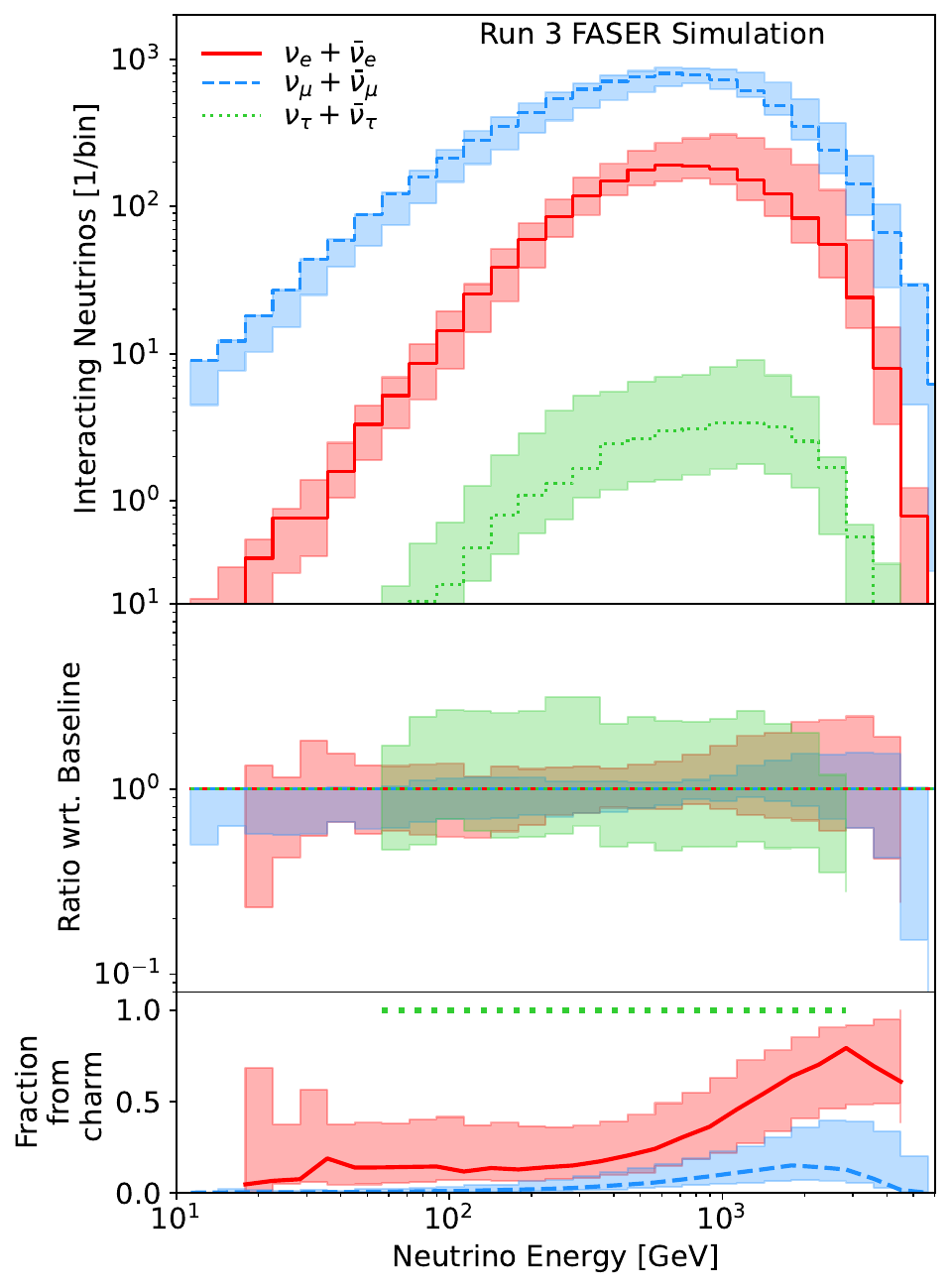}
\caption{The energy spectra of neutrinos interacting in FASER$\nu$ at LHC Run 3 with an integrated luminosity of $250~\ifb$ for electron neutrinos (solid red), muon neutrinos (dashed blue), and tau neutrinos (dotted green). For each neutrino species, the central prediction is determined by the decay of light (charm) hadrons as predicted by \texttt{EPOS-LHC} (\texttt{POWHEG+Pythia 8.3}) with the interaction cross section provided by \texttt{GENIE}.  The shaded bands are the uncertainties due to the flux and do not include cross section uncertainties. The upper panel shows the energy spectra. The central panel shows the same data normalized to the central prediction. The lower panel shows the fraction of neutrinos produced in charm hadron decays.}
\label{fig:fasernu_spectra}
\end{figure}

\begin{figure}[tbph]
\centering
\includegraphics[width = 0.99\textwidth]{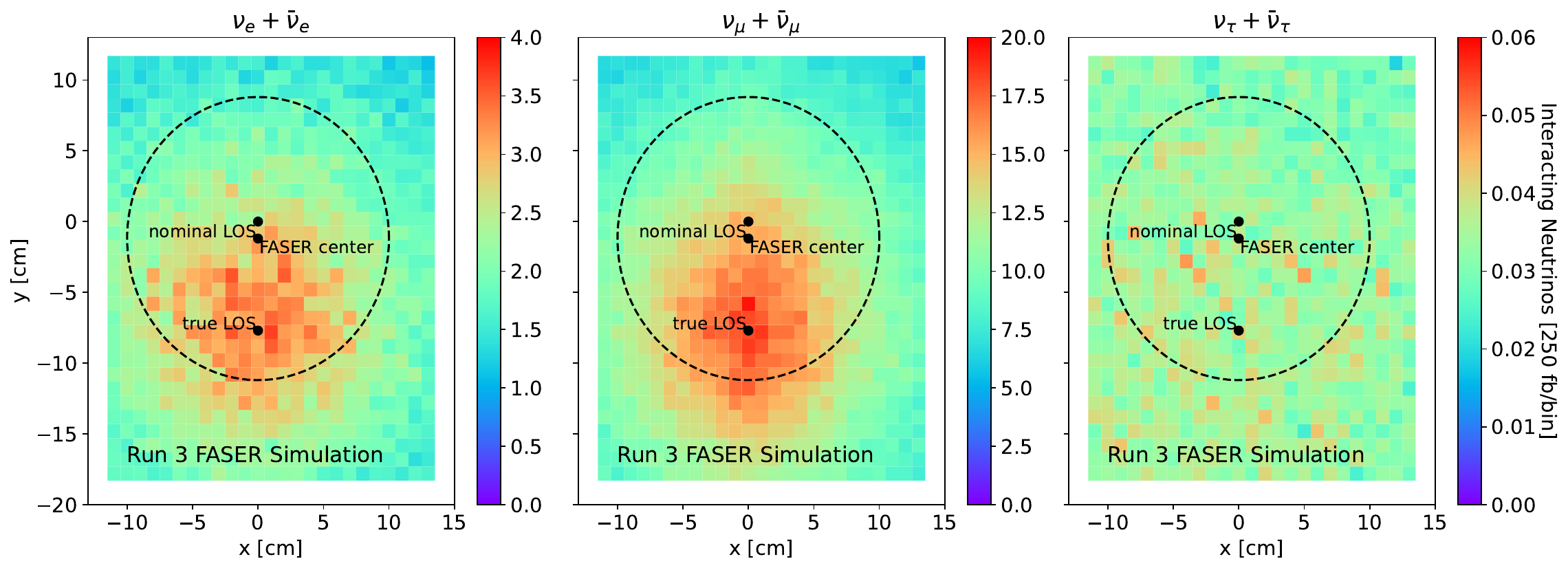}
\caption{The distribution in the transverse plane in nominal coordinates of neutrinos interacting in LHC Run 3 with an integrated luminosity of $250~\ifb$ in FASER$\nu$ for electron (left), muon (middle), and tau (right) neutrinos, with light and charm hadron production modeled with \texttt{EPOS-LHC} and \texttt{POWHEG+Pythia 8.3}, respectively. The dashed black circle indicates the area covered by the FASER spectrometer.}  
\label{fig:2D_fasernu}
\end{figure}

The spatial distribution in the transverse plane of neutrinos interacting in FASER$\nu$ at LHC Run 3 with a total integrated luminosity of $250~\ifb$ is shown in \cref{fig:2D_fasernu}. The three panels show results for the three neutrino species, $\nu_e$, $\nu_{\mu}$, and $\nu_{\tau}$, from left to right. \texttt{EPOS-LHC} and \texttt{POWHEG+Pythia~8.3} for light and charm hadron production are used, respectively. Here, the integration has been performed over all energies.  At relatively low energies, muon neutrinos, produced primarily in pion decays, are more collimated near the LOS than the other flavors; electron neutrinos, dominantly produced in kaon decays, are less collimated; and tau neutrinos, produced in charm hadron decays, are the least collimated. At $\sim \tev$ energies, muon neutrinos are produced dominantly in kaon decays, electron neutrinos have significant contributions from both kaon and charm hadron decay, and tau neutrinos are again produced only in charm hadron decays, leading to the same relative ordering of collimation.

In \cref{fig:2D_fasernu_5}, the distribution of interacting neutrinos is shown in the $(x, \text{Energy})$ plane, where $x$ is the horizontal spatial coordinate; this plot would look similar for the vertical component. Here a detector is assumed with the same material as FASER$\nu$, centered on the true LOS, but with dimensions $1~\m \times 1~\m\times 1~\m$, where the larger transverse extension is chosen to show the distribution over a larger range.  As in \cref{fig:2D_fasernu}, muon neutrinos are more collimated than electron neutrinos, which are more collimated than tau neutrinos, but we also see that the highest-energy neutrinos of each species are focused along the LOS.  

\begin{figure}[tbp]
\centering
\includegraphics[width = 0.99\textwidth]{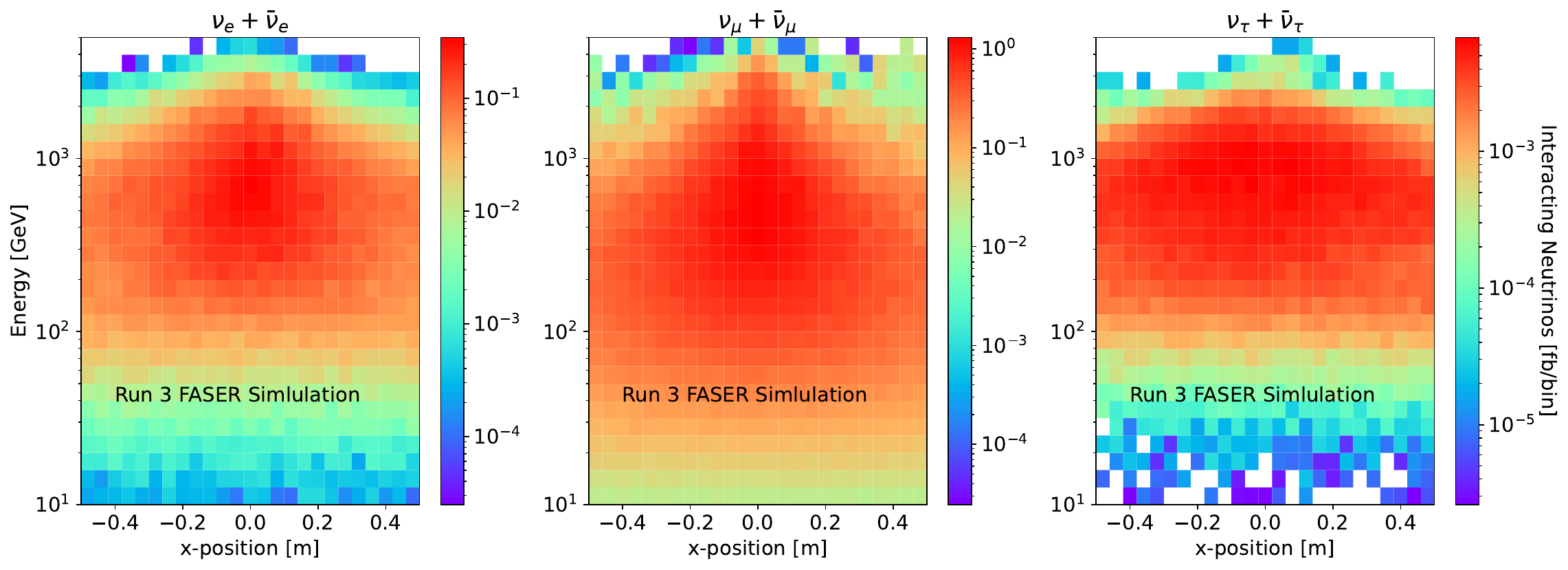}
\caption{The rate (in fb per bin) of neutrinos interacting in a $1~\m \times 1~\m\times 1~\m$ FASER$\nu$-like detector in the $(x, \text{Energy})$ plane in LHC Run 3.  The detector is centered on the true LOS, so the LOS is at $x=0$. Results are shown for electron (left), muon (middle), and tau (right) neutrinos with light and charm hadron production modeled with \texttt{EPOS-LHC} and \texttt{POWHEG+Pythia 8.3}, respectively. }
  \label{fig:2D_fasernu_5}
\end{figure}

In \cref{fig:event_distribution} the binned event rate is shown for $\nu_{e}$, $\nu_{\mu}$, and $\nu_{\tau}$ for Run 3 (upper panels) and Run 3 + Run 4 (lower panels). For each energy bin, the colors show the composition in terms of the parent hadron.  Also shown for each bin is the statistical uncertainty, defined as $\sqrt{N_{\rm bin}}$. For $\nu_e$ the event rate is dominated by kaon decays at low energy, with charm decays becoming comparable at higher energies. Hyperon decays also provide a non-negligible contribution to the $\nu_e$ event rate. For $\nu_{\mu}$ the event rate is dominated by pion decays at low energies and kaon decays at high energies, with charm decays providing a sub-leading component. Charm hadrons are the only hadrons that produce $\nu_{\tau}$ at FASER$\nu$. In the absence of sizeable systematic uncertainties, we find that Run 3 will provide sufficient number of events to distinguish the contributions of the various parent hadrons to the neutrino flux detected at FASER.

\begin{figure}[tbp]
\centering
\includegraphics[width=0.32\textwidth]{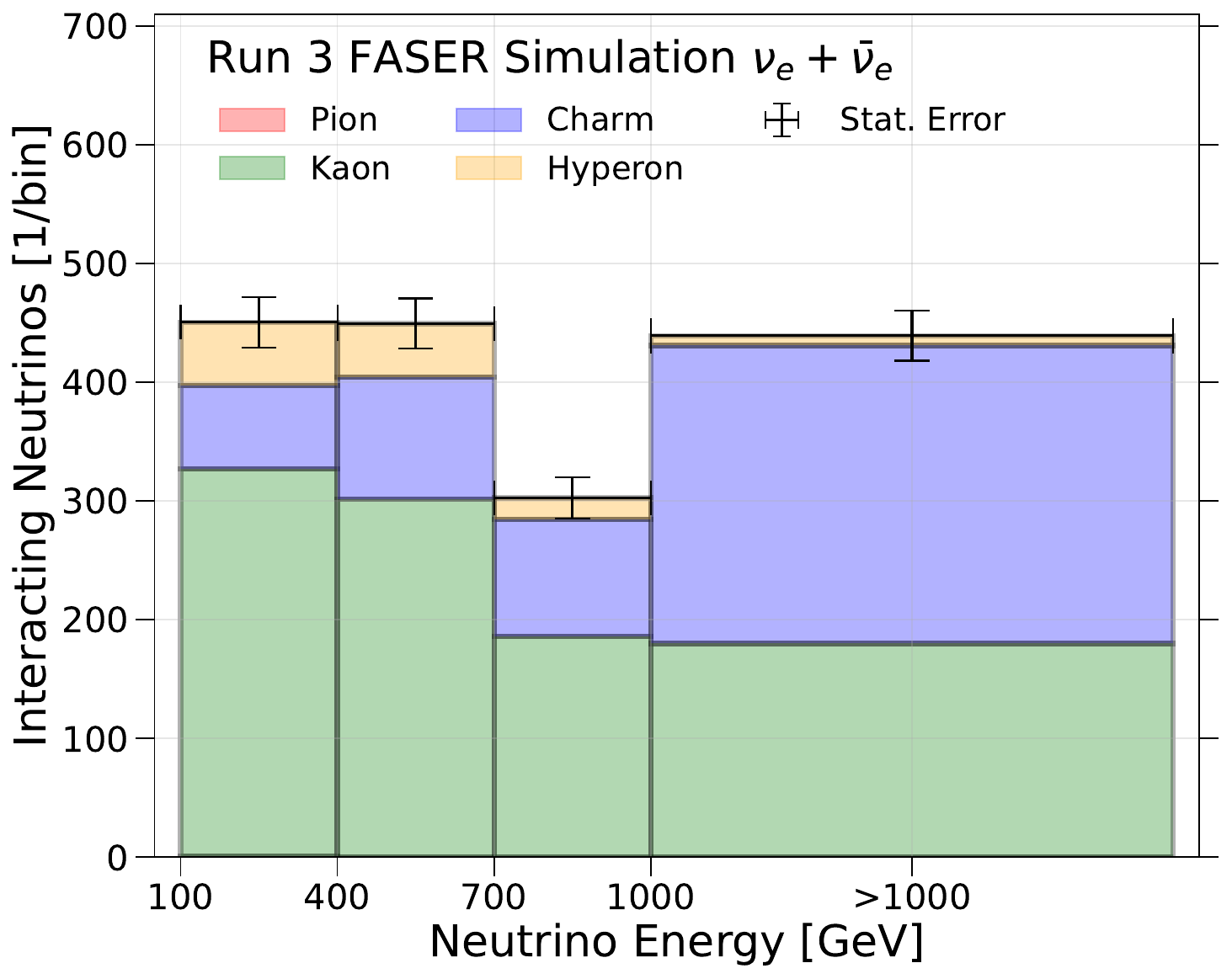}
\includegraphics[width=0.32\textwidth]{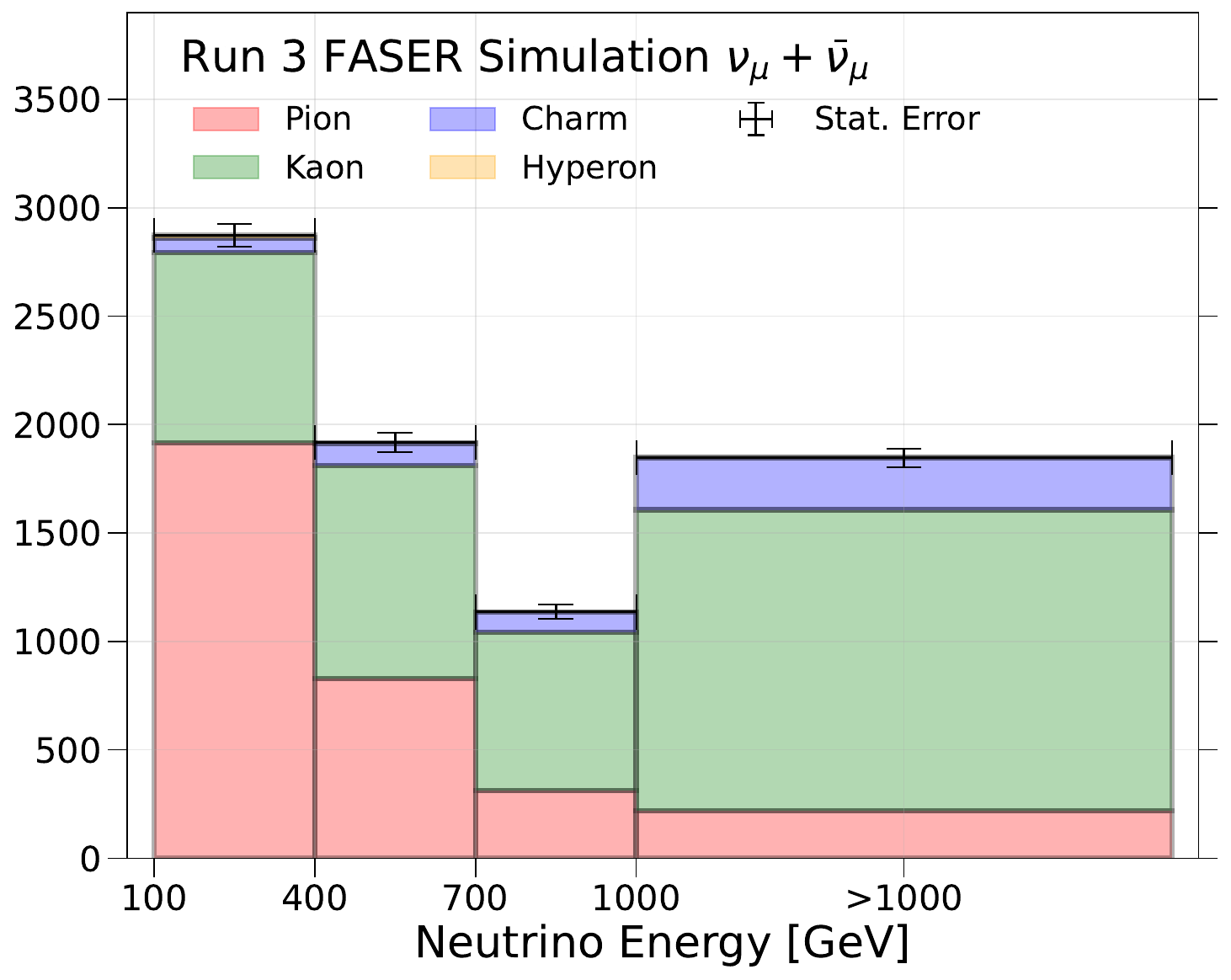}
\includegraphics[width=0.32\textwidth]{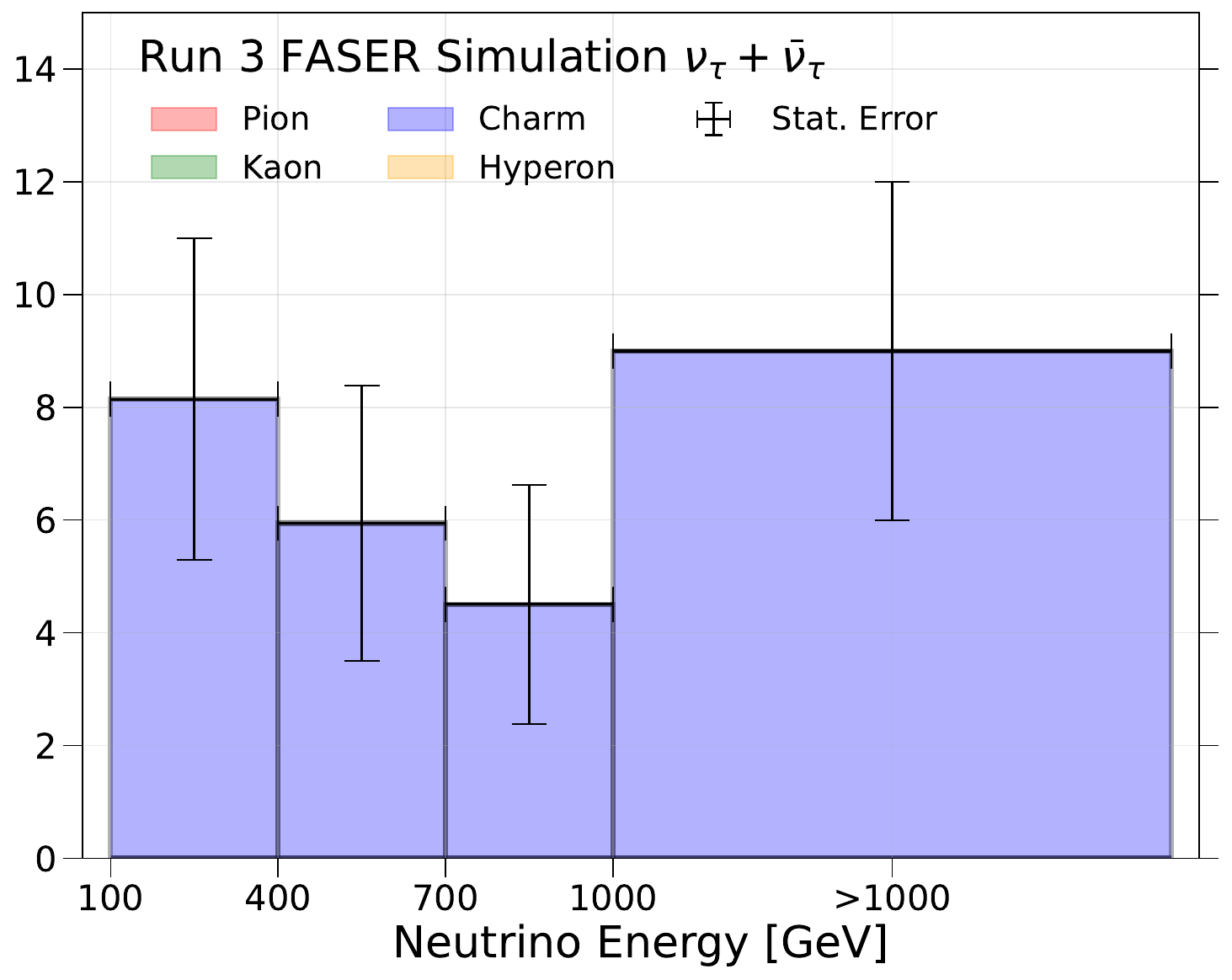}
\includegraphics[width=0.32\textwidth]{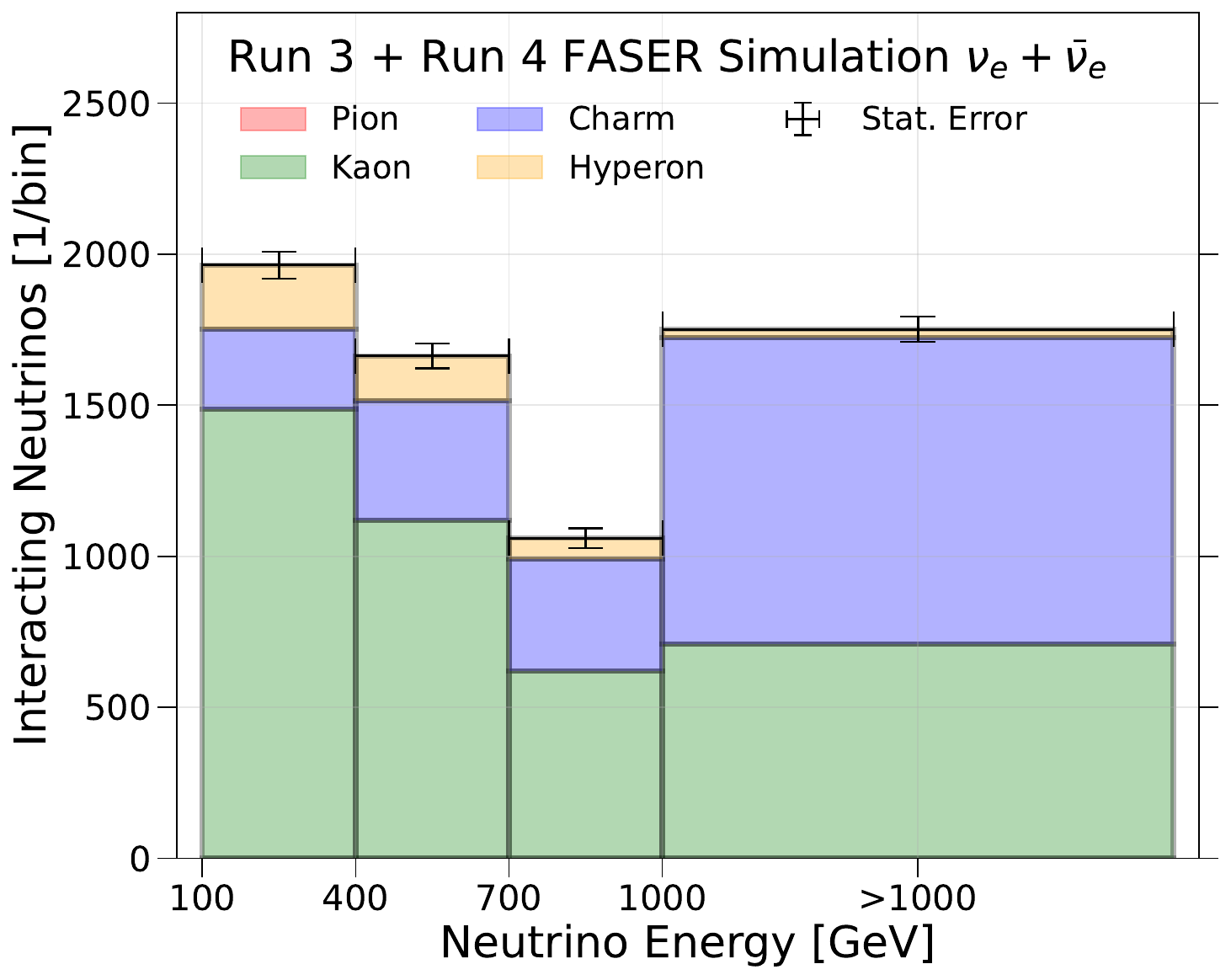}
\includegraphics[width=0.32\textwidth]{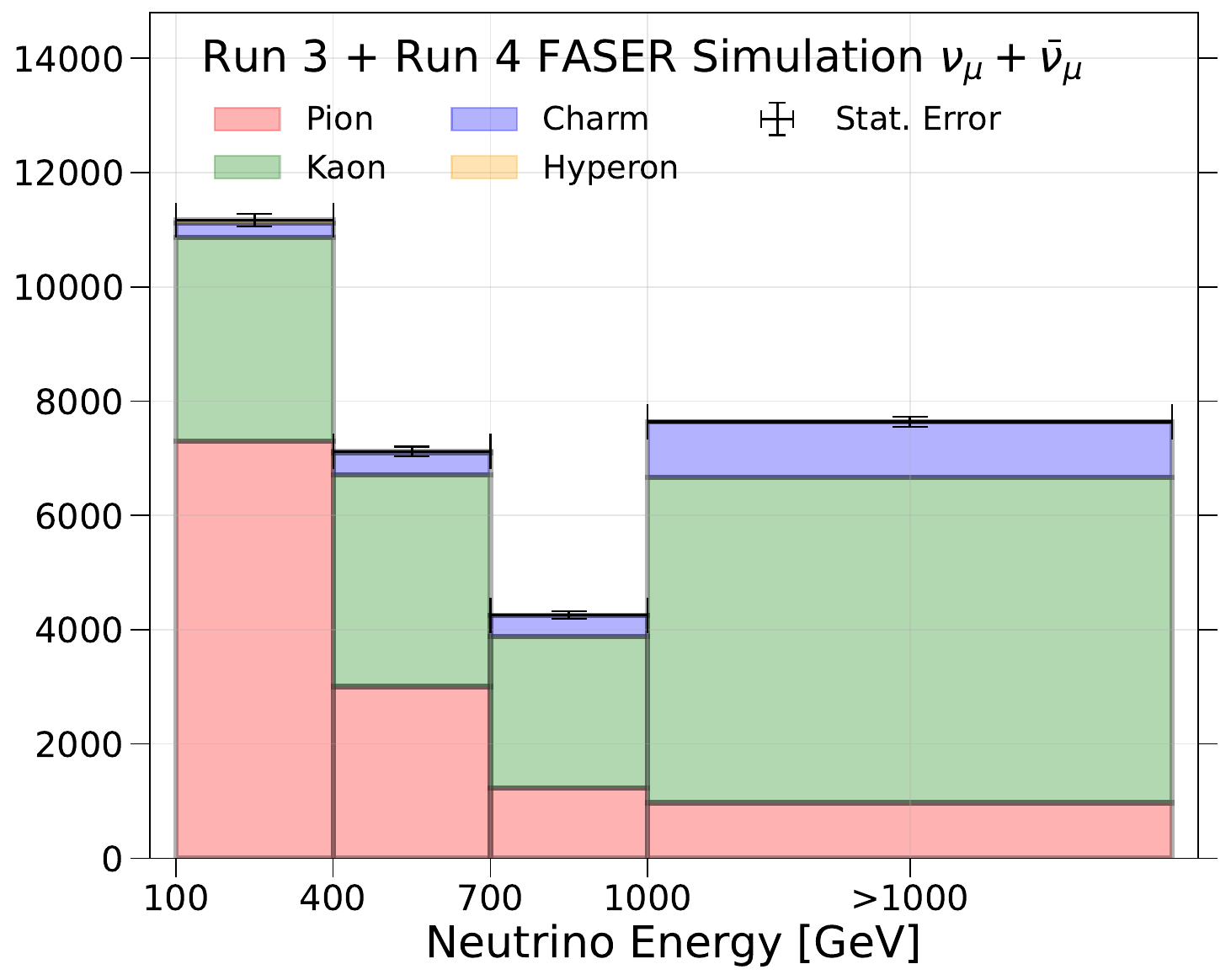}
\includegraphics[width=0.32\textwidth]{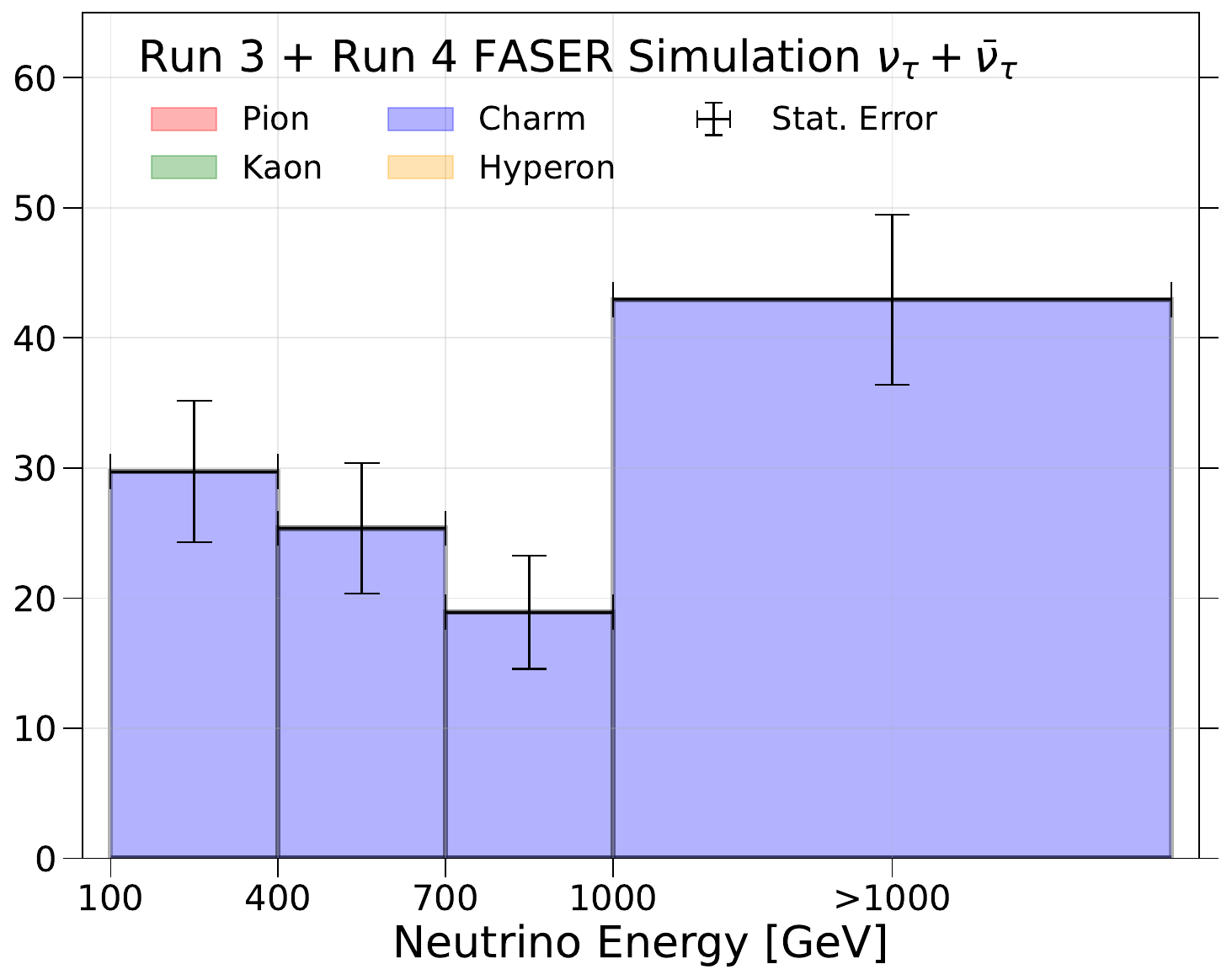}
\caption{Binned energy spectra for electron (left), muon (center), and tau (left) neutrinos interacting in FASER$\nu$ at LHC Run 3 with a total integrated luminosity of 250 $\ifb$ (upper panels) and at LHC Run 3 + Run 4 with a total integrated luminosity 930 $\ifb$ (lower panels). For each bin, the neutrinos are separated by their parent hadrons: pions (red), kaons (green), charm hadrons (blue), and hyperons (yellow). Hadron production and decay of light (charm) hadrons are modeled by \texttt{EPOS-LHC} (\texttt{POWHEG+Pythia~8.3}). Also shown are statistical errors per bin, defined as $\sqrt{N_{\rm bin}}$, which demonstrate that Run 3 measurements will have sufficient statistics to be sensitive to components of neutrinos from the different parent hadrons.}
\label{fig:event_distribution}
\end{figure} 

In \cref{fig:event_distribution_electronic} the energy distribution of muon neutrinos with CC interactions in FASER$\nu$ is shown. In Ref.~\cite{FASER:2023zcr} the electronic detector components of FASER were used to track the outgoing muon in CC $\nu_{\mu}$ interactions. By measuring the curvature of the muon in the magnetic field of FASER, the charge of the muon was identified for muons with energy below a TeV, allowing one to distinguish $\nu_\mu$ and $\bar{\nu}_\mu$ interactions. In \cref{fig:event_distribution_electronic}, the energy distributions are therefore separated into $\nu_{\mu}$ and $\bar{\nu}_{\mu}$ for energies below 1 TeV.  Above ${\cal O}(1~{\rm TeV})$ the charge measurement becomes unreliable, so $\nu_{\mu}$ and $\bar{\nu}_{\mu}$ are grouped together. Here, only neutrinos that interact within the portion of FASER$\nu$ that is within the aperture of the FASER spectrometer are included,  so that the forward-going muon can be measured. Also included is a $20\%$ efficiency for this signal, which is typical for this measurement at FASER~\cite{FASER:2023zcr}. By measuring the muon's charge, measurements by the electronic detector components of FASER will be able to distinguish $\nu_{\mu}$ from $\bar{\nu}_{\mu}$, providing a further probe of forward hadron production.

\begin{figure}[tbp]
\centering
\includegraphics[width=0.49\textwidth]{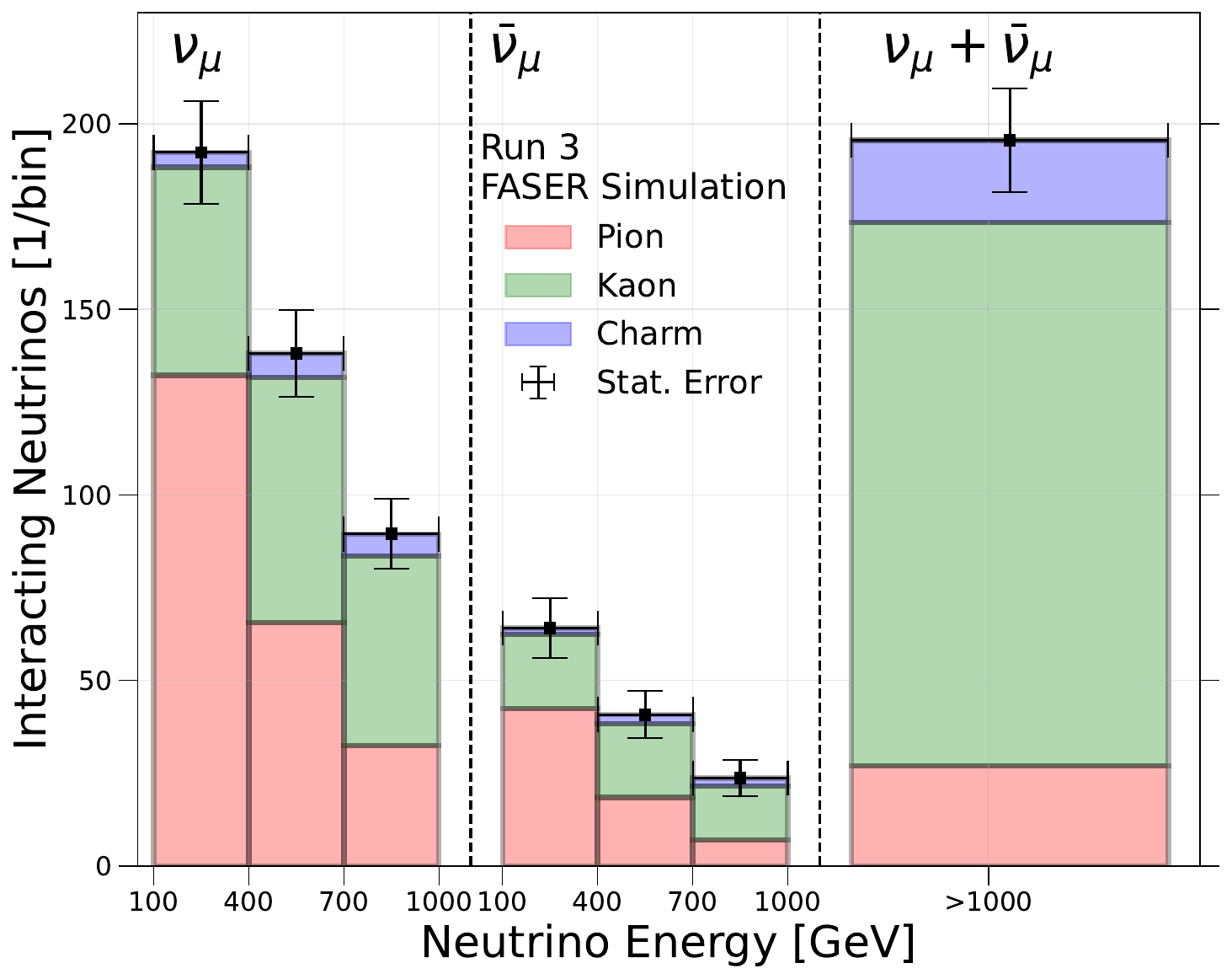}
\includegraphics[width=0.49\textwidth]{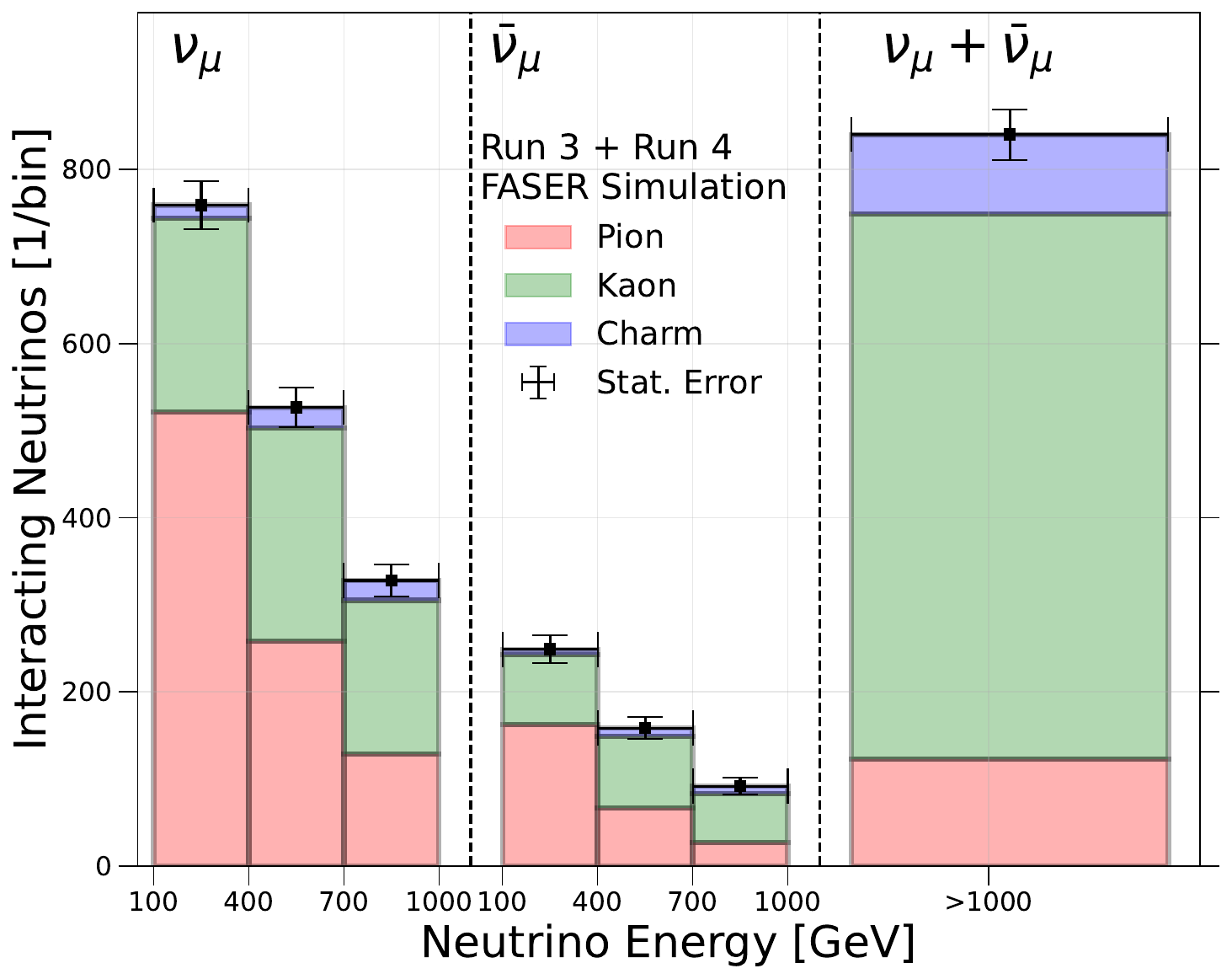}
\caption{The energy spectra of muon neutrinos and anti-neutrinos detected through their CC interactions, using only the electronic detector components of FASER in LHC Run 3 with a total integrated luminosity of $250~\ifb$ (left) and in LHC Run 3 + Run 4 with a total integrated luminosity of $930~\ifb$ (right). For each bin, the neutrinos are separated by their parent hadrons, as indicated.  FASER's magnets enable charge identification of the outgoing muons, which allows $\nu_{\mu}$ and $\bar{\nu}_{\mu}$ separation for energies below a TeV. }
\label{fig:event_distribution_electronic}
\end{figure}

\section{Conclusions}
\label{sec:conclusion}

The recent discovery of collider neutrinos at FASER has opened up the new field of TeV laboratory neutrinos. Their large fluxes and relatively large interaction cross sections imply large event rates, even with relatively small detectors.  To fully realize the potential of these neutrino events for both SM and BSM physics~\cite{Anchordoqui:2021ghd, Feng:2022inv}, it is necessary to have accurate predictions of the forward neutrino spectrum.

The forward neutrino flux at the LHC is dominantly produced by the decays of pions, kaons, hyperons, and charm hadrons. As the forward region is relatively unprobed, there are sizeable uncertainties in the fluxes of these parent hadrons.  For neutrinos from light hadron decays, the central value is taken to be the predictions of \texttt{EPOS-LHC}~\cite{Pierog:2013ria}, with the associated uncertainty given by the spread in predictions from the event generators \texttt{EPOS-LHC}~\cite{Pierog:2013ria}, \texttt{Sibyll}~\cite{Riehn:2019jet}, \texttt{QGSJET}~\cite{Ostapchenko:2010vb}, and \pythiaF~\cite{Fieg:2023kld}.  For neutrinos from charm hadron decays, the central value is taken to be the results of \texttt{POWHEG+Pythia 8.3}~\cite{Alioli:2010xd, Bierlich:2022pfr}, with the uncertainty given by the variation resulting from varying the factorization and renormalization scales~\cite{Buonocore:2023kna}.

The corresponding neutrino fluxes are obtained using the dedicated fast neutrino flux simulation of Ref.~\cite{Kling:2021gos}, which has been updated for the Run 3 and Run 4 beam configurations. To produce the spectra of CC neutrino interactions in FASER$\nu$, the Bodek-Yang model~\cite{Bodek:2002vp, Bodek:2004pc, Bodek:2010km} implemented in \texttt{GENIE}~\cite{GENIE:2021wox} is used, which agrees with more recent cross section calculations for TeV neutrinos, \texttt{NNSFv}~\cite{Candido:2023utz} and \texttt{CKMT+PCAC-NT}~\cite{Jeong:2023hwe}, to within $\lesssim 6\%$ over the range of energies of interest.

The expected neutrino event rates are presented in \cref{tab:interactions}.  These include results for LHC Run 3 with an expected integrated luminosity of $250~\ifb$.  In addition, given the recent approval of FASER for LHC Run 4~\cite{Run4LOI}, neutrino event distributions are simulated for LHC Run 4 with an expected total integrated luminosity of $680~\ifb$. The central values of the expected neutrino event rates for $\nu_e$, $\nu_{\mu}$, and $\nu_{\tau}$ are 1700, 8500, and 30 in LHC Run 3 and 4900, 25000, and 90 in LHC Run 4.  Such event rates imply percent-level statistical uncertainties for electron and muon neutrino studies.  For tau neutrinos, the number of events that will be observed at FASER in the coming years will greatly enhance the number that have been observed to date.  

Last, in \cref{fig:event_distribution,fig:event_distribution_electronic} results are presented for both Run 3 and Run 3 + Run 4 for the energy spectra of neutrinos interacting in FASER$\nu$, decomposed into components based on the parent hadron species.  It can be seen that statistical uncertainties will be small enough that FASER$\nu$ will be sensitive not only to the leading contributions, but also to sub-leading contributions. Provided experimental systematic uncertainties are not dominant, these results imply promising prospects for studying very high energy neutrinos, forward hadron production, and their many related topics with FASER in the coming years at LHC Run 3 and Run 4.

\acknowledgements

We thank Luca Buonocore, Anatoli Fedynitch, Alfonso Garcia, Yu Seon Jeong, Hallsie Reno, Luca Rottoli, and Dennis Soldin for useful discussions. 
We thank the technical and administrative staff members at all FASER institutions for their contributions to the success of the FASER project. 
This work was supported in part by Heising-Simons Foundation Grant Nos.~2018-1135, 2019-1179, and 2020-1840, Simons Foundation Grant No.~623683, U.S.~National Science Foundation Grant Nos.~PHY-2111427, PHY-2110929, and PHY-2110648, JSPS KAKENHI Grants Nos.~JP19H01909, JP20K23373, JP20H01919, JP20K04004, and JP21H00082, BMBF Grant No.~05H20PDRC1, DFG EXC 2121 Quantum Universe Grant No. 390833306, ERC Consolidator Grant No.~101002690, Royal Society Grant No.~URF\textbackslash R1\textbackslash 201519, UK Science and Technology Funding Councils Grant No.~ST/ T505870/1, the National Natural Science Foundation of China, Tsinghua University Initiative Scientific Research Program, and the Swiss National Science Foundation.

\appendix*

\section{Comparison of Forward Charm Production Models}
\label{sec:charmmodels}

As evident in \cref{fig:event_distribution}, forward charm hadron production plays an important role in determining neutrino event rates at FASER, contributing significantly to $\nu_e$ and $\nu_{\mu}$ rates, and providing essentially all of the $\nu_{\tau}$ rate.  In this work, the central value for forward charm production is taken to be the results of \texttt{POWHEG+Pythia 8.3}~\cite{Alioli:2010xd, Bierlich:2022pfr}, with the uncertainty given by varying the factorization and renormalization scales~\cite{Buonocore:2023kna}.

Here we compare the results of this prescription to the results from other generators.  In \cref{fig:charmcomparison}, we show results from \texttt{POWHEG+Pythia 8.3}, along with results from \texttt{SIBYLL 2.3d}~\cite{Riehn:2019jet}, \texttt{Pythia~8.3}~\cite{Sjostrand:2014zea}, and \texttt{DPMJET~3.2019.1}~\cite{DPMJET}.  As noted by the author~\cite{Anatoli}, \texttt{DPMJET} was never validated for charm production and is not intended to be used for forward charm production.  However, \texttt{DPMJET} is used by FLUKA~\cite{Battistoni:2015epi}, a widely used framework for propagating particles through the LHC infrastructure and estimating event rates in forward detectors, and so it is instructive to include it here for comparison.  

In \cref{fig:charmcomparison} (left), results are shown for charm hadron production for pseudorapidities $\eta$ in the range 2 to 4.5, along with data from LHCb. \texttt{POWHEG+Pythia 8.3}, \texttt{SIBYLL}, and \texttt{Pythia} give comparable predictions, and the variation in these generators is well characterized by varying the factorization and renormalization scales in \texttt{POWHEG+Pythia 8.3}, as prescribed in this work.  These results are also consistent with LHCb data.  In contrast, \texttt{DPMJET} deviates from the other three generators and predicts charm hadron rates that are inconsistent with the data.  This inconsistency may be attributed to a number of aspects of the \texttt{DPMJET} model: (i) the assumption of massless charm quarks in calculating the underlying matrix element for $gg \to c \bar{c}$; (ii) the use of CT14LO PDFs that may overestimate the charm quark content; and (iii) a $k$ factor ($\sim 2$) that, although not unreasonably large in this context, significantly enhances the rate.

The results of these generators for the spectrum of electron neutrinos detected at FASER$\nu$ is shown in \cref{fig:charmcomparison} (right). Once again, \texttt{POWHEG+Pythia 8.3}, \texttt{SIBYLL}, and \texttt{Pythia} give comparable predictions, and the variation between them is fairly well characterized by the uncertainty prescription for \texttt{POWHEG+Pythia 8.3}. However, \texttt{DPMJET} predicts much larger event rates, differing from the other prescriptions by a factor of 10 or even larger at the highest energies.  Until the discrepancy with LHCb data is understood and resolved, projections for neutrino event rates originating from \texttt{DPMJET} charm hadron predictions cannot be considered on a solid footing and may be overestimating the flux by as much as an order of magnitude.

\begin{figure}[tbp]
\includegraphics[width=0.494\textwidth]{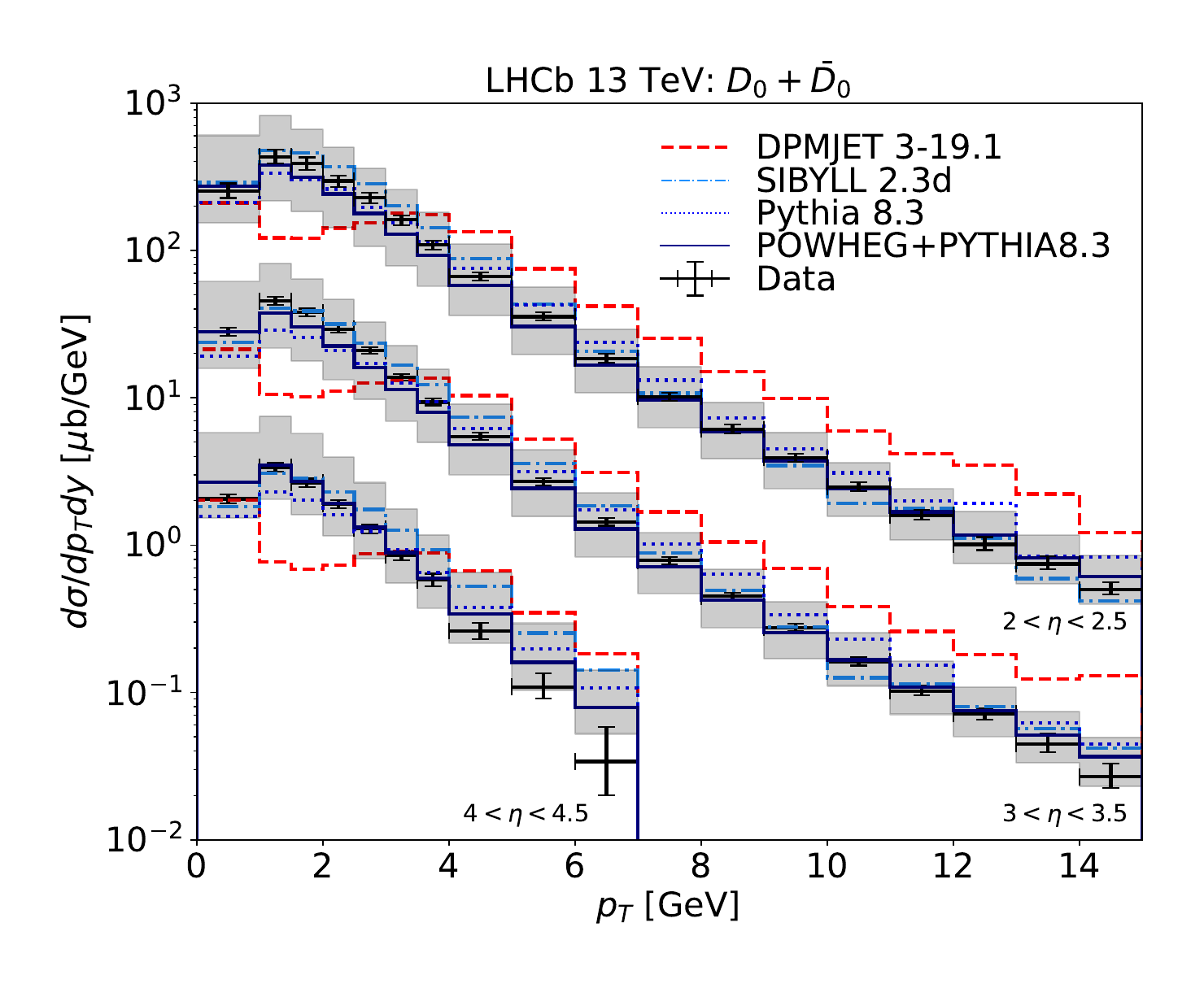}
\includegraphics[width=0.482\textwidth]{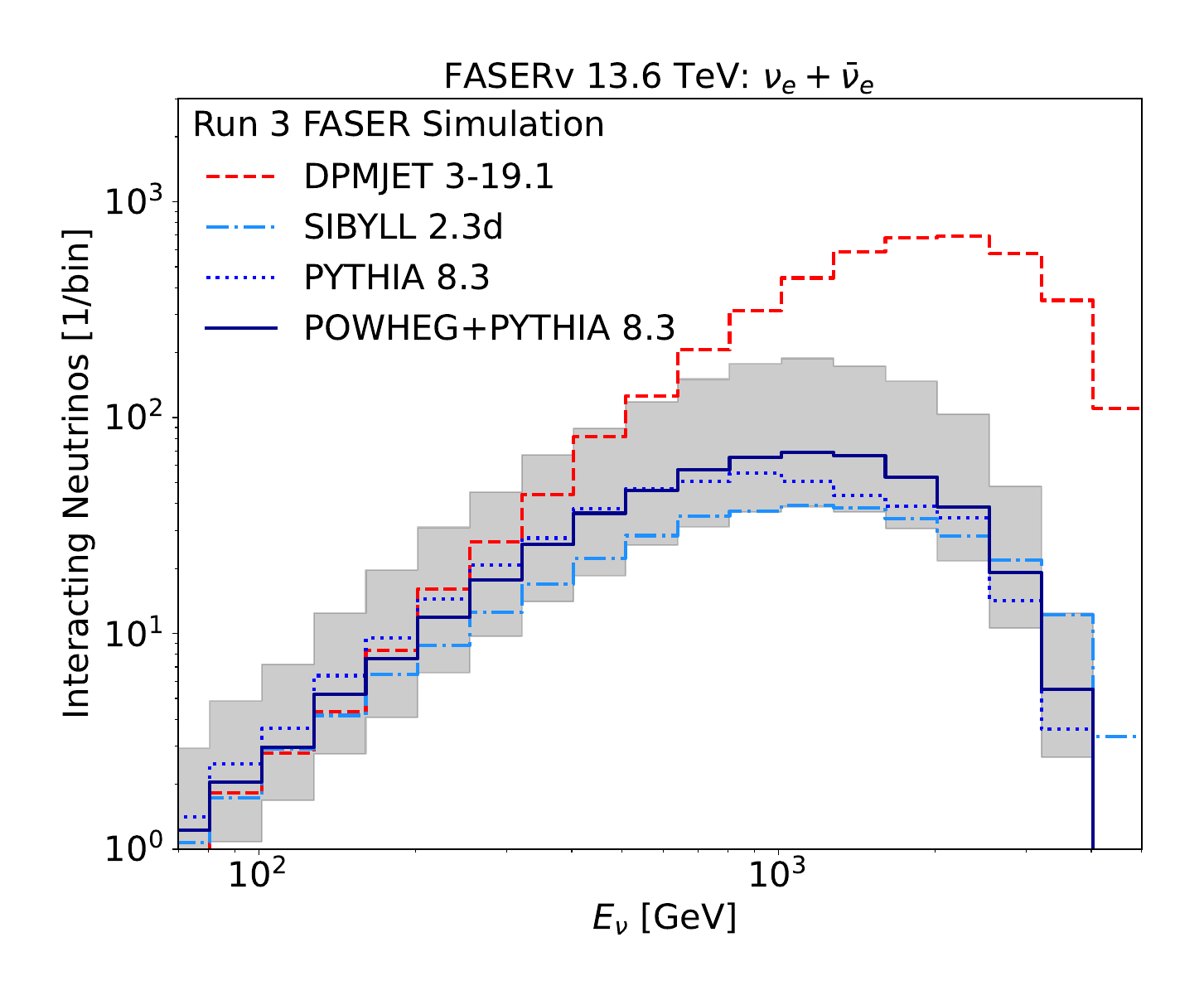}
\caption{Comparison of Forward Charm Production Models.  Left: Comparison of estimates of charm hadron fluxes from \texttt{POWHEG+Pythia 8.3}, \texttt{SIBYLL}, \texttt{Pythia}, and \texttt{DPMJET} for pseudorapidities $2 < \eta < 4.5$, along with data from LHCb for $pp$ collisions at center-of-mass energy $\sqrt{s} = 13~\tev$~\cite{LHCb:2015swx}.  The spectra for the $3<\eta<3.5$ and $4<\eta<4.5$ pseudorapidity regions are rescaled by factors of 10 and 100, respectively, for display purposes.  Right: For the same models, the predicted energy spectra of electron neutrinos from charm hadron decay that interact in FASER$\nu$ during Run 3 at $\sqrt{s} = 13.6~\tev$. The gray-shaded region in each panel is the uncertainty envelope for the \texttt{POWHEG+Pythia 8.3} prediction, as determined by varying the factorization and renormalization scales.
\label{fig:charmcomparison}
}
\end{figure}

\bibliography{references}

\providecommand{\href}[2]{#2}\begingroup\raggedright\begin{thebibliography}{10}

\bibitem{Feng:2017uoz}
J.~L. Feng, I.~Galon, F.~Kling, and S.~Trojanowski, ``{ForwArd Search ExpeRiment at the LHC},'' \href{http://dx.doi.org/10.1103/PhysRevD.97.035001}{{\em Phys. Rev. D} {\bf 97} (2018) no.~3, 035001}, \href{http://arxiv.org/abs/1708.09389}{{\tt arXiv:1708.09389 [hep-ph]}}.

\bibitem{FASER:2022hcn}
{\bf FASER} Collaboration, ``{The FASER Detector},'' \href{http://arxiv.org/abs/2207.11427}{{\tt arXiv:2207.11427 [physics.ins-det]}}.

\bibitem{FASER:2018bac}
{\bf FASER} Collaboration, ``{Technical Proposal for FASER: ForwArd Search ExpeRiment at the LHC},'' \href{http://arxiv.org/abs/1812.09139}{{\tt arXiv:1812.09139 [physics.ins-det]}}.

\bibitem{FASER:2020gpr}
{\bf FASER} Collaboration, ``{Technical Proposal: FASERnu},'' \href{http://arxiv.org/abs/2001.03073}{{\tt arXiv:2001.03073 [physics.ins-det]}}.

\bibitem{FASER:2021cpr}
{\bf FASER} Collaboration, ``{The trigger and data acquisition system of the FASER experiment},'' \href{http://dx.doi.org/10.1088/1748-0221/16/12/P12028}{{\em JINST} {\bf 16} (2021) no.~12, P12028}, \href{http://arxiv.org/abs/2110.15186}{{\tt arXiv:2110.15186 [physics.ins-det]}}.

\bibitem{FASER:2021ljd}
{\bf FASER} Collaboration, ``{The tracking detector of the FASER experiment},'' \href{http://dx.doi.org/10.1016/j.nima.2022.166825}{{\em Nucl. Instrum. Meth. A} {\bf 1034} (2022)  166825}, \href{http://arxiv.org/abs/2112.01116}{{\tt arXiv:2112.01116 [physics.ins-det]}}.

\bibitem{FASER:2023zcr}
{\bf FASER} Collaboration, ``{First Direct Observation of Collider Neutrinos with FASER at the LHC},'' \href{http://dx.doi.org/10.1103/PhysRevLett.131.031801}{{\em Phys. Rev. Lett.} {\bf 131} (2023) no.~3, 031801}, \href{http://arxiv.org/abs/2303.14185}{{\tt arXiv:2303.14185 [hep-ex]}}.

\bibitem{CERN-FASER-CONF-2023-002}
{\bf FASER} Collaboration, ``{Observation of high-energy electron neutrino interactions with FASER's emulsion detector at the LHC},'' tech. rep., CERN, Geneva, 2023.
\newblock \url{https://cds.cern.ch/record/2868284}.

\bibitem{FASER:2023tle}
{\bf FASER} Collaboration, ``{Search for Dark Photons with the FASER detector at the LHC},'' \href{http://dx.doi.org/10.1016/j.physletb.2023.138378}{{\em Phys. Lett. B} {\bf 848} (2024)  138378}, \href{http://arxiv.org/abs/2308.05587}{{\tt arXiv:2308.05587 [hep-ex]}}.

\bibitem{SNDLHC:2023pun}
{\bf SND@LHC} Collaboration, ``{Observation of Collider Muon Neutrinos with the SND@LHC Experiment},'' \href{http://dx.doi.org/10.1103/PhysRevLett.131.031802}{{\em Phys. Rev. Lett.} {\bf 131} (2023) no.~3, 031802}, \href{http://arxiv.org/abs/2305.09383}{{\tt arXiv:2305.09383 [hep-ex]}}.

\bibitem{Run4LOI}
{\bf FASER} Collaboration, ``{Request to run FASER in Run 4},'' Tech. Rep. CERN-LHCC-2023-009, CERN, Geneva, November, 2023.
\newblock \url{https://cds.cern.ch/record/2882503}.

\bibitem{Kling:2023tgr}
F.~Kling, T.~M\"akel\"a, and S.~Trojanowski, ``{Investigating the fluxes and physics potential of LHC neutrino experiments},'' \href{http://dx.doi.org/10.1103/PhysRevD.108.095020}{{\em Phys. Rev. D} {\bf 108} (2023) no.~9, 095020}, \href{http://arxiv.org/abs/2309.10417}{{\tt arXiv:2309.10417 [hep-ph]}}.

\bibitem{Cruz-Martinez:2023sdv}
J.~M. Cruz-Martinez, M.~Fieg, T.~Giani, P.~Krack, T.~M\"akel\"a, T.~Rabemananjara, and J.~Rojo, ``{The LHC as a Neutrino-Ion Collider},'' \href{http://arxiv.org/abs/2309.09581}{{\tt arXiv:2309.09581 [hep-ph]}}.

\bibitem{Anchordoqui:2022fpn}
L.~A. Anchordoqui, C.~G. Canal, F.~Kling, S.~J. Sciutto, and J.~F. Soriano, ``{An explanation of the muon puzzle of ultrahigh-energy cosmic rays and the role of the Forward Physics Facility for model improvement},'' \href{http://dx.doi.org/10.1016/j.jheap.2022.03.004}{{\em JHEAp} {\bf 34} (2022)  19--32}, \href{http://arxiv.org/abs/2202.03095}{{\tt arXiv:2202.03095 [hep-ph]}}.

\bibitem{Anchordoqui:2021ghd}
L.~A. Anchordoqui {\em et al.}, ``{The Forward Physics Facility: Sites, experiments, and physics potential},'' \href{http://dx.doi.org/10.1016/j.physrep.2022.04.004}{{\em Phys. Rept.} {\bf 968} (2022)  1--50}, \href{http://arxiv.org/abs/2109.10905}{{\tt arXiv:2109.10905 [hep-ph]}}.

\bibitem{Feng:2022inv}
J.~L. Feng {\em et al.}, ``{The Forward Physics Facility at the High-Luminosity LHC},'' \href{http://dx.doi.org/10.1088/1361-6471/ac865e}{{\em J. Phys. G} {\bf 50} (2023) no.~3, 030501}, \href{http://arxiv.org/abs/2203.05090}{{\tt arXiv:2203.05090 [hep-ex]}}.

\bibitem{Kling:2021gos}
F.~Kling and L.~J. Nevay, ``{Forward neutrino fluxes at the LHC},'' \href{http://dx.doi.org/10.1103/PhysRevD.104.113008}{{\em Phys. Rev. D} {\bf 104} (2021) no.~11, 113008}, \href{http://arxiv.org/abs/2105.08270}{{\tt arXiv:2105.08270 [hep-ph]}}.

\bibitem{Bodek:2002vp}
A.~Bodek and U.~K. Yang, ``{Modeling deep inelastic cross-sections in the few GeV region},'' \href{http://dx.doi.org/10.1016/S0920-5632(02)01755-3}{{\em Nucl. Phys. B Proc. Suppl.} {\bf 112} (2002)  70--76}, \href{http://arxiv.org/abs/hep-ex/0203009}{{\tt arXiv:hep-ex/0203009}}.

\bibitem{Bodek:2004pc}
A.~Bodek, I.~Park, and U.-k. Yang, ``{Improved low Q**2 model for neutrino and electron nucleon cross sections in few GeV region},'' \href{http://dx.doi.org/10.1016/j.nuclphysbps.2004.11.208}{{\em Nucl. Phys. B Proc. Suppl.} {\bf 139} (2005)  113--118}, \href{http://arxiv.org/abs/hep-ph/0411202}{{\tt arXiv:hep-ph/0411202}}.

\bibitem{Bodek:2010km}
A.~Bodek and U.-k. Yang, ``{Axial and Vector Structure Functions for Electron- and Neutrino- Nucleon Scattering Cross Sections at all $Q^2$ using Effective Leading order Parton Distribution Functions},'' \href{http://arxiv.org/abs/1011.6592}{{\tt arXiv:1011.6592 [hep-ph]}}.

\bibitem{Buckley:2010ar}
A.~Buckley, J.~Butterworth, D.~Grellscheid, H.~Hoeth, L.~Lonnblad, J.~Monk, H.~Schulz, and F.~Siegert, ``{Rivet user manual},'' \href{http://dx.doi.org/10.1016/j.cpc.2013.05.021}{{\em Comput. Phys. Commun.} {\bf 184} (2013)  2803--2819}, \href{http://arxiv.org/abs/1003.0694}{{\tt arXiv:1003.0694 [hep-ph]}}.

\bibitem{Bierlich:2019rhm}
C.~Bierlich {\em et al.}, ``{Robust Independent Validation of Experiment and Theory: Rivet version 3},'' \href{http://dx.doi.org/10.21468/SciPostPhys.8.2.026}{{\em SciPost Phys.} {\bf 8} (2020)  026}, \href{http://arxiv.org/abs/1912.05451}{{\tt arXiv:1912.05451 [hep-ph]}}.

\bibitem{Nevay:2018zhp}
L.~J. Nevay {\em et al.}, ``{BDSIM: An accelerator tracking code with particle\textendash{}matter interactions},'' \href{http://dx.doi.org/10.1016/j.cpc.2020.107200}{{\em Comput. Phys. Commun.} {\bf 252} (2020)  107200}, \href{http://arxiv.org/abs/1808.10745}{{\tt arXiv:1808.10745 [physics.comp-ph]}}.

\bibitem{Agostinelli:2002hh}
S.~Agostinelli {\em et al.}, ``{\textsc{Geant4} -- a simulation toolkit},''
\href{http://dx.doi.org/10.1016/S0168-9002(03)01368-8}{{\em Nucl. Instrum. Meth. A} {\bf 506} (2003)  250}.

\bibitem{ZurbanoFernandez:2020cco}
I.~Zurbano~Fernandez {\em et al.}, \href{http://dx.doi.org/10.23731/CYRM-2020-0010}{``{High-Luminosity Large Hadron Collider (HL-LHC): Technical design report},''} vol.~10/2020.
\newblock 12, 2020.

\bibitem{MAD-X}
``{MAD --- Methodical Accelerator Design},''. \url{https://madx.web.cern.ch/madx}.

\bibitem{TomasGarcia:2803611}
R.~Tomas~Garcia {\em et al.}, ``{HL-LHC Run 4 proton operational scenario},'' tech. rep., CERN, Geneva, 2022.
\newblock \url{https://cds.cern.ch/record/2803611}.

\bibitem{Albrecht:2021cxw}
J.~Albrecht {\em et al.}, ``{The Muon Puzzle in cosmic-ray induced air showers and its connection to the Large Hadron Collider},'' \href{http://dx.doi.org/10.1007/s10509-022-04054-5}{{\em Astrophys. Space Sci.} {\bf 367} (2022) no.~3, 27}, \href{http://arxiv.org/abs/2105.06148}{{\tt arXiv:2105.06148 [astro-ph.HE]}}.

\bibitem{Pierog:2013ria}
T.~Pierog, I.~Karpenko, J.~M. Katzy, E.~Yatsenko, and K.~Werner, ``{EPOS LHC: Test of collective hadronization with data measured at the CERN Large Hadron Collider},'' \href{http://dx.doi.org/10.1103/PhysRevC.92.034906}{{\em Phys. Rev. C} {\bf 92} (2015) no.~3, 034906}, \href{http://arxiv.org/abs/1306.0121}{{\tt arXiv:1306.0121 [hep-ph]}}.

\bibitem{Riehn:2019jet}
F.~Riehn, R.~Engel, A.~Fedynitch, T.~K. Gaisser, and T.~Stanev, ``{Hadronic interaction model Sibyll 2.3d and extensive air showers},'' \href{http://dx.doi.org/10.1103/PhysRevD.102.063002}{{\em Phys. Rev. D} {\bf 102} (2020) no.~6, 063002}, \href{http://arxiv.org/abs/1912.03300}{{\tt arXiv:1912.03300 [hep-ph]}}.

\bibitem{Ostapchenko:2010vb}
S.~Ostapchenko, ``{Monte Carlo treatment of hadronic interactions in enhanced Pomeron scheme: I. QGSJET-II model},'' \href{http://dx.doi.org/10.1103/PhysRevD.83.014018}{{\em Phys. Rev. D} {\bf 83} (2011)  014018}, \href{http://arxiv.org/abs/1010.1869}{{\tt arXiv:1010.1869 [hep-ph]}}.

\bibitem{Sjostrand:2014zea}
T.~Sj\"ostrand, S.~Ask, J.~R. Christiansen, R.~Corke, N.~Desai, P.~Ilten, S.~Mrenna, S.~Prestel, C.~O. Rasmussen, and P.~Z. Skands, ``{An introduction to PYTHIA 8.2},'' \href{http://dx.doi.org/10.1016/j.cpc.2015.01.024}{{\em Comput. Phys. Commun.} {\bf 191} (2015)  159--177}, \href{http://arxiv.org/abs/1410.3012}{{\tt arXiv:1410.3012 [hep-ph]}}.

\bibitem{Bierlich:2022pfr}
C.~Bierlich {\em et al.}, ``{A comprehensive guide to the physics and usage of PYTHIA 8.3},'' \href{http://arxiv.org/abs/2203.11601}{{\tt arXiv:2203.11601 [hep-ph]}}.

\bibitem{Fieg:2023kld}
M.~Fieg, F.~Kling, H.~Schulz, and T.~Sj\"ostrand, ``{Tuning pythia for forward physics experiments},'' \href{http://dx.doi.org/10.1103/PhysRevD.109.016010}{{\em Phys. Rev. D} {\bf 109} (2024) no.~1, 016010}, \href{http://arxiv.org/abs/2309.08604}{{\tt arXiv:2309.08604 [hep-ph]}}.

\bibitem{LHCf:2008lfy}
{\bf LHCf} Collaboration, ``{The LHCf detector at the CERN Large Hadron Collider},'' \href{http://dx.doi.org/10.1088/1748-0221/3/08/S08006}{{\em JINST} {\bf 3} (2008)  S08006}.

\bibitem{LHCf:2017fnw}
{\bf LHCf} Collaboration, ``{Measurement of forward photon production cross-section in proton-proton collisions at $\sqrt{s}$ = 13 TeV with the LHCf detector},'' \href{http://dx.doi.org/10.1016/j.physletb.2017.12.050}{{\em Phys. Lett. B} {\bf 780} (2018)  233--239}, \href{http://arxiv.org/abs/1703.07678}{{\tt arXiv:1703.07678 [hep-ex]}}.

\bibitem{Piparo:2023yam}
{\bf LHCf} Collaboration, ``{Measurement of the forward \ensuremath{\eta} meson production rate in p-p collisions at $ \sqrt{\textrm{s}} $ = 13 TeV with the LHCf-Arm2 detector},'' \href{http://dx.doi.org/10.1007/JHEP10(2023)169}{{\em JHEP} {\bf 10} (2023)  169}, \href{http://arxiv.org/abs/2305.06633}{{\tt arXiv:2305.06633 [hep-ex]}}.

\bibitem{LHCf:2018gbv}
{\bf LHCf} Collaboration, ``{Measurement of inclusive forward neutron production cross section in proton-proton collisions at $ \sqrt{s}=13 $ TeV with the LHCf Arm2 detector},'' \href{http://dx.doi.org/10.1007/JHEP11(2018)073}{{\em JHEP} {\bf 11} (2018)  073}, \href{http://arxiv.org/abs/1808.09877}{{\tt arXiv:1808.09877 [hep-ex]}}.

\bibitem{LHCf:2015rcj}
{\bf LHCf} Collaboration, ``{Measurements of longitudinal and transverse momentum distributions for neutral pions in the forward-rapidity region with the LHCf detector},'' \href{http://dx.doi.org/10.1103/PhysRevD.94.032007}{{\em Phys. Rev. D} {\bf 94} (2016) no.~3, 032007}, \href{http://arxiv.org/abs/1507.08764}{{\tt arXiv:1507.08764 [hep-ex]}}.

\bibitem{GENIE:2021wox}
{\bf GENIE} Collaboration, ``{Hadronization model tuning in genie v3},'' \href{http://dx.doi.org/10.1103/PhysRevD.105.012009}{{\em Phys. Rev. D} {\bf 105} (2022) no.~1, 012009}, \href{http://arxiv.org/abs/2106.05884}{{\tt arXiv:2106.05884 [hep-ph]}}.

\bibitem{DPMJET}
A.~Fedynitch, ``{DPMJET},''. \url{https://github.com/DPMJET}.

\bibitem{Roesler:2000he}
S.~Roesler, R.~Engel, and J.~Ranft, \href{http://dx.doi.org/10.1007/978-3-642-18211-2_166}{``{The Monte Carlo event generator DPMJET-III},''} in {\em {International Conference on Advanced Monte Carlo for Radiation Physics, Particle Transport Simulation and Applications (MC 2000)}}, pp.~1033--1038.
\newblock 12, 2000.
\newblock \href{http://arxiv.org/abs/hep-ph/0012252}{{\tt arXiv:hep-ph/0012252}}.

\bibitem{Bai:2020ukz}
W.~Bai, M.~Diwan, M.~V. Garzelli, Y.~S. Jeong, and M.~H. Reno, ``{Far-forward neutrinos at the Large Hadron Collider},'' \href{http://dx.doi.org/10.1007/JHEP06(2020)032}{{\em JHEP} {\bf 06} (2020)  032}, \href{http://arxiv.org/abs/2002.03012}{{\tt arXiv:2002.03012 [hep-ph]}}.

\bibitem{Maciula:2022lzk}
R.~Maciula and A.~Szczurek, ``{Far-forward production of charm mesons and neutrinos at forward physics facilities at the LHC and the intrinsic charm in the proton},'' \href{http://dx.doi.org/10.1103/PhysRevD.107.034002}{{\em Phys. Rev. D} {\bf 107} (2023) no.~3, 034002}, \href{http://arxiv.org/abs/2210.08890}{{\tt arXiv:2210.08890 [hep-ph]}}.

\bibitem{Bhattacharya:2023zei}
A.~Bhattacharya, F.~Kling, I.~Sarcevic, and A.~M. Stasto, ``{Forward neutrinos from charm at the Large Hadron Collider},'' \href{http://dx.doi.org/10.1103/PhysRevD.109.014040}{{\em Phys. Rev. D} {\bf 109} (2024) no.~1, 014040}, \href{http://arxiv.org/abs/2306.01578}{{\tt arXiv:2306.01578 [hep-ph]}}.

\bibitem{Buonocore:2023kna}
L.~Buonocore, F.~Kling, L.~Rottoli, and J.~Sominka, ``{Predictions for neutrinos and new physics from forward heavy hadron production at the LHC},'' \href{http://dx.doi.org/10.1140/epjc/s10052-024-12726-5}{{\em Eur. Phys. J. C} {\bf 84} (2024) no.~4, 363}, \href{http://arxiv.org/abs/2309.12793}{{\tt arXiv:2309.12793 [hep-ph]}}.

\bibitem{Nason:2004rx}
P.~Nason, ``{A New method for combining NLO QCD with shower Monte Carlo algorithms},'' \href{http://dx.doi.org/10.1088/1126-6708/2004/11/040}{{\em JHEP} {\bf 11} (2004)  040}, \href{http://arxiv.org/abs/hep-ph/0409146}{{\tt arXiv:hep-ph/0409146}}.

\bibitem{Frixione:2007vw}
S.~Frixione, P.~Nason, and C.~Oleari, ``{Matching NLO QCD computations with Parton Shower simulations: the POWHEG method},'' \href{http://dx.doi.org/10.1088/1126-6708/2007/11/070}{{\em JHEP} {\bf 11} (2007)  070}, \href{http://arxiv.org/abs/0709.2092}{{\tt arXiv:0709.2092 [hep-ph]}}.

\bibitem{Alioli:2010xd}
S.~Alioli, P.~Nason, C.~Oleari, and E.~Re, ``{A general framework for implementing NLO calculations in shower Monte Carlo programs: the POWHEG BOX},'' \href{http://dx.doi.org/10.1007/JHEP06(2010)043}{{\em JHEP} {\bf 06} (2010)  043}, \href{http://arxiv.org/abs/1002.2581}{{\tt arXiv:1002.2581 [hep-ph]}}.

\bibitem{Ball:2017otu}
R.~D. Ball, V.~Bertone, M.~Bonvini, S.~Marzani, J.~Rojo, and L.~Rottoli, ``{Parton distributions with small-x resummation: evidence for BFKL dynamics in HERA data},'' \href{http://dx.doi.org/10.1140/epjc/s10052-018-5774-4}{{\em Eur. Phys. J. C} {\bf 78} (2018) no.~4, 321}, \href{http://arxiv.org/abs/1710.05935}{{\tt arXiv:1710.05935 [hep-ph]}}.

\bibitem{Bertone:2018dse}
V.~Bertone, R.~Gauld, and J.~Rojo, ``{Neutrino Telescopes as QCD Microscopes},'' \href{http://dx.doi.org/10.1007/JHEP01(2019)217}{{\em JHEP} {\bf 01} (2019)  217}, \href{http://arxiv.org/abs/1808.02034}{{\tt arXiv:1808.02034 [hep-ph]}}.

\bibitem{Candido:2023utz}
A.~Candido, A.~Garcia, G.~Magni, T.~Rabemananjara, J.~Rojo, and R.~Stegeman, ``{Neutrino Structure Functions from GeV to EeV Energies},'' \href{http://dx.doi.org/10.1007/JHEP05(2023)149}{{\em JHEP} {\bf 05} (2023)  149}, \href{http://arxiv.org/abs/2302.08527}{{\tt arXiv:2302.08527 [hep-ph]}}.

\bibitem{Jeong:2023hwe}
Y.~S. Jeong and M.~H. Reno, ``{Neutrino cross sections: Interface of shallow- and deep-inelastic scattering for collider neutrinos},'' \href{http://dx.doi.org/10.1103/PhysRevD.108.113010}{{\em Phys. Rev. D} {\bf 108} (2023) no.~11, 113010}, \href{http://arxiv.org/abs/2307.09241}{{\tt arXiv:2307.09241 [hep-ph]}}.

\bibitem{Gluck:1998xa}
M.~Gl\"uck, E.~Reya, and A.~Vogt, ``{Dynamical parton distributions revisited},'' \href{http://dx.doi.org/10.1007/s100520050289}{{\em Eur. Phys. J. C} {\bf 5} (1998)  461--470}, \href{http://arxiv.org/abs/hep-ph/9806404}{{\tt arXiv:hep-ph/9806404}}.

\bibitem{Cooper-Sarkar:2011jtt}
A.~Cooper-Sarkar, P.~Mertsch, and S.~Sarkar, ``{The high energy neutrino cross-section in the Standard Model and its uncertainty},'' \href{http://dx.doi.org/10.1007/JHEP08(2011)042}{{\em JHEP} {\bf 08} (2011)  042}, \href{http://arxiv.org/abs/1106.3723}{{\tt arXiv:1106.3723 [hep-ph]}}.

\bibitem{Garcia:2020jwr}
A.~Garcia, R.~Gauld, A.~Heijboer, and J.~Rojo, ``{Complete predictions for high-energy neutrino propagation in matter},'' \href{http://dx.doi.org/10.1088/1475-7516/2020/09/025}{{\em JCAP} {\bf 09} (2020)  025}, \href{http://arxiv.org/abs/2004.04756}{{\tt arXiv:2004.04756 [hep-ph]}}.

\bibitem{Xie:2023suk}
K.~Xie, J.~Gao, T.~J. Hobbs, D.~R. Stump, and C.~P. Yuan, ``{High-energy neutrino deeply inelastic scattering cross sections from 100 GeV to 1000 EeV},'' \href{http://arxiv.org/abs/2303.13607}{{\tt arXiv:2303.13607 [hep-ph]}}.

\bibitem{Anatoli}
A.~Fedynitch. Private communication.

\bibitem{Battistoni:2015epi}
G.~Battistoni {\em et al.}, ``{Overview of the FLUKA code},'' \href{http://dx.doi.org/10.1016/j.anucene.2014.11.007}{{\em Annals Nucl. Energy} {\bf 82} (2015)  10--18}.

\bibitem{LHCb:2015swx}
{\bf LHCb} Collaboration, ``{Measurements of prompt charm production cross-sections in $pp$ collisions at $ \sqrt{s}=13 $ TeV},'' \href{http://dx.doi.org/10.1007/JHEP03(2016)159}{{\em JHEP} {\bf 03} (2016)  159}, \href{http://arxiv.org/abs/1510.01707}{{\tt arXiv:1510.01707 [hep-ex]}}. [Erratum: JHEP 09, 013 (2016), Erratum: JHEP 05, 074 (2017)].

\end{thebibliography}\endgroup

\end{document}